\documentclass[aps,pra,preprint,groupedaddress,amsmath,amssymb,showpacs]{revtex4-1}
\usepackage{graphicx,amsmath,amssymb,epsfig}
\usepackage{dcolumn}
\usepackage{bm}
\usepackage{color}
\allowdisplaybreaks
\begin{document}
\title{Analogue of double-$\Lambda$-type atomic medium and vector dromions in a plasmonic metamaterial}
\author{Qi Zhang$^{1}$, Zhengyang Bai$^{1}$, and Guoxiang Huang$^{1,2, }$\footnote{gxhuang@phy.ecnu.edu.cn} }
\affiliation{$^1$State Key Laboratory of Precision Spectroscopy,
                 East China Normal University, Shanghai 200062, China\\
             $^2$NYU-ECNU Joint Institute of Physics at NYU-Shanghai, Shanghai 200062, China
             }
\date{\today}

\begin{abstract}

We consider an array of the meta-atom consisting of two cut-wires and a split-ring resonator interacting with an electromagnetic field with two polarization components. We prove that such metamaterial system can be taken as a classical analogue of an atomic medium with a double-$\Lambda$-type four-level configuration coupled with four laser fields, exhibits an effect of plasmon induced transparency (PIT), and displays a similar behavior of atomic four-wave mixing (FWM). We demonstrate that when nonlinear varactors are mounted onto the gaps of the split-ring resonators the system can acquire giant second- and third-order Kerr nonlinearities via the PIT and a longwave-shortwave interaction. We also demonstrate that the system supports high-dimensional vector plasmonic dromions [i.e. (2+1)-dimensional plasmonic solitons with two polarization components, each of which has a coupling between a longwave and a shortwave], which have very low generation power and are robust during propagation. Our work gives not only a plasmonic analogue of the FWM in coherent atomic systems but also provides the possibility for obtaining new type of nonlinear polaritons in plasmonic metamaterials.

\end{abstract}

\maketitle

%



\section{Introduction}

Electromagnetically induced transparency (EIT), a very intriguing phenomenon occurring in atomic gases, has been intensively investigated due to its interesting physical properties and promising practical applications. The basic mechanism of EIT is the existence of a destructive quantum interference effect between two pathways of atomic transitions induced by a control laser field, through which the absorption of a probe laser field can be largely cancelled~\cite{Fleischhauer}.

In recent years, there are tremendous efforts seeking for the classical analogue of EIT in solid systems, including coupled resonators~\cite{Alzar,Spring,Harden}, electric circuits~\cite{Alzar,Harden,Souza}, optomechanical devices~\cite{Weis,Kronwald}, whispering-gallery-mode microresonators~\cite{Peng}, and various metamaterials~\cite{Zhangs,Papa,Tassin,Gu,han,Chen0,Dong,NLiu,Liu,TNakanishi,Sun,Chen,Mats,NakanishiPRApp,bai,Bai2,Bai3}. Especially, the plasmonic analogue of EIT, called plasmon-induced transparency (PIT)~\cite{Zhangs,Papa,Tassin}, has become a very important platform for exploring EIT-like physical properties of plasmonic polaritons and for designing new types of metematerials.

Similar to EIT, PIT is resulted from a destructive interference effect between wideband bright and narrowband dark modes in artificial atoms (called meta-atoms). A typical character of PIT is the opening of a deep transparency window within broadband absorption spectrum, together with a steep dispersion and greatly reduced group velocity of plasmonic polaritons.
PIT metamaterials can work in different frequency regions (including micro~\cite{Papa} and terahertz~\cite{Tassin,Gu,Dong} waves, infrared and visible radiations~\cite{Zhangs,NLiu,han}), and may be used to design novel, chip-scale plasmonic devices (including highly sensitive sensors~\cite{Dong,Chen0}, optical buffers~\cite{Gu,Liu}, and ultrafast optical switches~\cite{Gu}, etc.) in which the radiation damping can be significantly eliminated, very intriguing for practical applications.

However, the PIT in plasmonic metamaterials reported up to now~\cite{Zhangs,Papa,Tassin,Gu,han,Chen0,Dong,NLiu,Liu,TNakanishi,Sun,Chen,Mats,NakanishiPRApp,bai,Bai2,Bai3} is only for the classical analogue of the simplest atomic EIT, i.e. the one occurring in a coherent three-level atom gas resonantly interacting with two laser fields. We know that atoms possess many (energy) levels, quantum interference effect may occur in atomic systems with level number larger than three and the number of laser fields larger than two~\cite{Fleischhauer}.  In fact, in the past two decades the EIT has been extended into the atomic systems with various multi-level configurations, such as four-level systems of double $\Lambda$-type~\cite{Lukin,Korsunsky1,Korsunsky2,Merriam,DengPRA1,Kang,Ying,DengPRA3,YingSoliton,Kang1,IteYu}, tripod-type~\cite{Petrosyan,Rebic,Beck}, Y-type~\cite{Joshi,GaoJY,Kha,LiuYM},
five-level systems of M-type~\cite{Ott,Mat,Rebic1,HangC}, and six-level systems of double-tripod-type~\cite{Rus,Lee}, etc.
Thus it is natural to ask the question: Is it possible to get a classical analogue of the atomic EIT with the level number
more than three in a metamaterial?

In this article, we give a positive answer for the above question. The metamaterial we consider is assumed to be an array of meta-atoms~[see Fig.~\ref{Fig2}(a)], i.e. the unit cells consisting of two cut-wires (CWs) and a split-ring resonator (SRR)~[see Fig.~\ref{Fig2}(b)], interacting with an electromagnetic (EM) field with two polarization components.
We show that such plasmonic metamaterial system may be taken as a classical analogue of an atomic medium with a double-$\Lambda$-type four-level configuration coupled with four (two probe and two control) laser fields [see Fig.~\ref{Fig1}(a)\,], exhibits an effect of PIT and displays a similar behavior of atomic four-wave mixing (FWM).

Based on this classical analogue, we further show that, if nonlinear varactors are mounted onto the gaps of the SRRs, the system can acquire giant second- and third-order Kerr nonlinearities many orders of magnitude larger than conventional nonlinear optical media. Using a method of multiple scales, we derive coupled envelope equations, which include dispersion, diffraction, and the Kerr nonlinearities and govern the evolution of the two polarization components of the EM field. We demonstrate that the system supports a new type of nonlinear plasmonic polaritons, i.e. high-dimensional {\it vector} plasmonic dromions, which are (2+1)-dimensional plasmonic solitons with two polarization components. Each polarization component has a coupling between a longwave and a shortwave, which have very low generation power and are robust during propagation. The results presented here not only gives a close metamaterial analogue of the EIT and FWM in multi-level atomic systems, useful to illustrate and find novel interference and nonlinear properties in solid systems, but also provides a way to obtain new type of plasmonic polaritons via suitable design of plasmonic metamaterials.

The main body of the article is arranged as follows. In Sec~\ref{sec2}, we give a simple introduction of the four-level atomic model allowing EIT and FWM, describe the metamaterial model, and show the similarity between the two models. The propagation of linear plasmonic polaritons in the metamaterial is discussed in detail. In Sec.~\ref{sec3}, we derive the coupled nonlinear envelope equations and present the vector plasmonic dromion solutions when the nonlinear varactors are mounted onto the gaps of the SRRs. Lastly, in Sec~\ref{sec4} we give a discussion and a summary of our work.  Details of some calculating results are given in five appendixes.

\section{EIT-based atomic FWM and its metamaterial analogue}\label{sec2}
\subsection{EIT-based FWM in a double-$\Lambda$-type four-level atomic system}

\begin{figure}
\includegraphics[width=0.8\columnwidth]{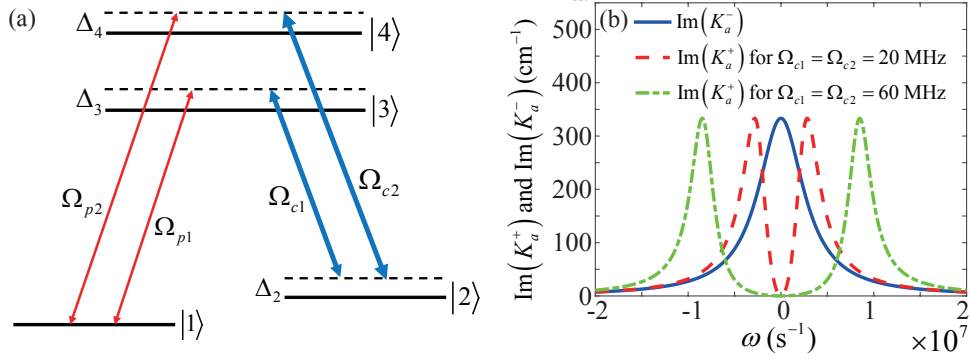}
\caption{(color online) (a) Double-$\Lambda$-type four-level atomic system with the atomic states $|j\rangle$ ($j=1,2,3,4$), coupled with two probe fields (with Rabi frequency $\Omega_{pn}$) and two strong control fields (with Rabi frequency $\Omega_{cn})$ $(n=1,2)$. $\Delta_{3}$, $\Delta_{2}$, and $\Delta_{4}$ are respectively the one, two, and three-photon detunings.
(b) ${\rm Im}(K_{a}^{+})$ [imaginary part of $K_{a}^{+}$] as a function of $\omega$ for $\Omega_{c1}=\Omega_{c2}=20\,{\rm MHz}$  (red dashed line) and $\Omega_{c1}=\Omega_{c2}=60\,{\rm MHz}$  (green dashed-dot line). EIT transparency window is opened near the central frequency of the probe fields (i.e. at $\omega=0$). The blue solid cure is ${\rm Im}(K_{a}^{-})$, which has always a large absorption peak at $\omega=0$ for arbitrary $\Omega_{c1}$ and $\Omega_{c2}$.
}
\label{Fig1}
\end{figure}

For a detailed comparison with the metamaterial model presented in the next subsection, we first give a brief introduction on a lifetime-broadened atomic gas with a double-$\Lambda$-type four-level configuration, shown in Fig.~\ref{Fig1}(a).
In this system, two weak probe laser fields with central angular frequencies $\omega_{p1}$ and $\omega_{p2}$ and wavevectors ${\bf k}_{p1}$ and ${\bf k}_{p2}$ drive respectively the transitions $|1\rangle\leftrightarrow|3\rangle$ and $|1\rangle\leftrightarrow|4\rangle$, and two strong control laser fields with central angular frequencies $\omega_{c1}$ and $\omega_{c2}$ and wavevectors ${\bf k}_{c1}$ and ${\bf k}_{c2}$ drive respectively the transitions $|2\rangle\leftrightarrow|3\rangle$ and $|2\rangle\leftrightarrow|4\rangle$. The total electric fields in this system is given by
${\bf E}={\bf e}_{p1}{\cal E}_{p1}\exp[i(k_{p1}z-\omega_{p1}t)]+{\bf e}_{p2}{\cal E}_{p2}\exp[i(k_{p2}z-\omega_{p2}t)]
+{\bf e}_{c1}{\cal E}_{c1}\exp[i(k_{c1}z-\omega_{c1}t)]+{\bf e}_{c2}{\cal E}_{c2}\exp[i(k_{c2}z-\omega_{c2}t)]+{\rm c.c.}$,
where ${\bf e}_{jn}$ and ${\cal E}_{jn}$  ($j=p,c;n=1,2$) are respectively the unit vector denoting the polarization direction and the envelope of the corresponding laser field. Note that for simplicity all the laser fields are assumed to be injected in the same (i.e. $z$) direction (which is also useful to suppress Doppler effect).  Under electric-dipole approximation and rotating-wave approximation (RWA), the Hamiltonian of the system in interaction picture reads
\begin{equation}
\hat {H}_{\rm int}=-{\hbar}\sum_{j=1}^{4}\Delta_{j}|j\rangle\langle j|
-\hbar\left[ \Omega_{p1}|3\rangle\langle1|+\Omega_{p2}|4\rangle\langle1|
+\Omega_{c1}|3\rangle\langle2|+\Omega_{c2}|4\rangle\langle2|+{\rm H.c.}\right],
\label{Ham}
\end{equation}
where $\Delta_{1}=0$; $\Delta_{3}=\omega_{p1}-(E_{3}-E_{1})/\hbar$, $\Delta_{2}=\omega_{p1}-\omega_{c1}-(E_{2}-E_{1})/\hbar$, and $\Delta_{4}=(\omega_{p1}-\omega_{c1}+\omega_{c2})-(E_{4}-E_{1})/\hbar$ are respectively one-,
two-, and three-photon detunings,
with $E_{l}$  the eigenenergy of the atomic state $|l\rangle$ ($l=1,2,3,4$);
$\Omega_{p1}=({\bf e}_{p1}\cdot {\bf p}_{31}){\cal E}_{p1}/\hbar$,
$\Omega_{p2}=({\bf e}_{p2}\cdot {\bf p}_{41}){\cal E}_{p2}/\hbar$,
$\Omega_{c1}=({\bf e}_{c1}\cdot {\bf p}_{32}){\cal E}_{c1}/\hbar$,
and $\Omega_{c2}=({\bf e}_{c2}\cdot {\bf p}_{42}){\cal E}_{c2}/\hbar$
are respectively the half Rabi frequencies of the probe and the control laser fields,
with ${\bf p}_{jl}$ the electric dipole moment related to the transition $|j\rangle\leftrightarrow|l\rangle$.
The Hamiltonian~(\ref{Ham}) allows three bright states and one dark state~\cite{note00}.
The dark state reads $|\psi_{\rm dark}\rangle=(\Omega_{c1}|1\rangle-\Omega_{p1}|2\rangle)/\sqrt{|\Omega_{p1}|^2+|\Omega_{c1}|^2}$,
%
%
which is a superposition of only the two lower states $|1\rangle$ and $|2\rangle$ and has a zero eigenvalue.
The condition yielding the dark state is~\cite{Korsunsky2}
\begin{equation}\label{Atoms Dark State}
\Omega_{p1}\Omega_{c2}-\Omega_{p2}\Omega_{c1}=0.
\end{equation}
%

The dynamics of the atoms is governed by the optical Bloch equation
$i\hbar\left(\partial/\partial t+\Gamma\right)\sigma=[{\hat H}_{\rm int},\sigma]$,
where $\sigma$ is a $4\times 4$ density matrix,
$\Gamma$ is a $4\times4$ decoherence (relaxation) matrix describing spontaneous emission and dephasing. The explicit expression of the Bloch equation is given in Appendix~\ref{app1}. We assume that initially the probe fields are absent, thus for substantially strong control fields the atoms are populated in the ground state $|1\rangle$. The solution of the Bloch equation reads $\sigma_{11}=1$ and all other $\sigma_{jl}$ are zero.

When the two weak probe fields are applied, the ground state $|1\rangle$ is not depleted much. In this case, the Bloch equation reduces to
\begin{subequations}\label{BEs}
\begin{align}
& \left(i\frac{\partial}{\partial t}+d_{31}\right)\sigma_{31}+\Omega_{c1}\sigma_{21}+\Omega_{p1}=0,\label{BEs1}\\
& \left(i\frac{\partial}{\partial t}+d_{41}\right)\sigma_{41}+\Omega_{c2}\sigma_{21}+\Omega_{p2}=0,\label{BEs2}\\
& \left(i\frac{\partial}{\partial t}+d_{21}\right)\sigma_{21}+\Omega_{c1}^{\ast}\sigma_{31}+\Omega_{c2}^{\ast}\sigma_{41}=0,\label{BEs3}
\end{align}
\end{subequations}
with $d_{j1}=\Delta_{j}+i\gamma_{j1}$ with $\gamma_{j1}=\Gamma_{1j}/2$ ($j=2,3,4$). Equations (\ref{BEs1})-(\ref{BEs3}) describe the dynamics of three coupled harmonic oscillators~\cite{note01}, where $\sigma_{31}$ and $\sigma_{41}$ are bright oscillators due to their direct coupling to the probe fields $\Omega_{p1}$ and $\Omega_{p2}$, but $\sigma_{21}$ is a dark oscillator because it has no direct coupling to any of the two probe fields.

The dynamics of the probe fields is governed by the Maxwell equation $\nabla^2{\bf E}-(1/c^2)\partial^2{\bf E}/{\partial t^2}=1/(\varepsilon_{0}c^2)\partial^2{\bf P}/{\partial t^2}$. Here the polarization intensity is given by ${\bf P}=N_{0}[\sigma_{31}e^{i(k_{p1}{ z}-\omega_{p1}t)}+\sigma_{41}e^{i(k_{p2}{z}-\omega_{p2}t)}+{\rm c.c.}]$, with $N_{0}$ the atomic density. Under a slowly-varying envelope approximation (SVEA), the Maxwell equation reduces to
\begin{subequations}\label{Atoms MEs}
\begin{align}
i\left(\frac{\partial}{\partial z}+\frac{1}{c}\frac{\partial}{\partial t}\right)\Omega_{p1}+\kappa_{13}\sigma_{31}=0,\\
i\left(\frac{\partial}{\partial z}+\frac{1}{c}\frac{\partial}{\partial t}\right)\Omega_{p2}+\kappa_{14}\sigma_{41}=0,
\end{align}
\end{subequations}
with $\kappa_{13}=N_{0}|{\bf e}_{p1}\cdot{\bf p}_{13}|^2\omega_{p1}/(2\hbar\varepsilon_{0}c)$ and $\kappa_{14}=N_{0}|{\bf e}_{p2}\cdot{\bf p}_{14}|^2\omega_{p2}/(2\hbar\varepsilon_{0}c)$. For simplicity, we assume the two control fields are strong enough and thus have no depletion during the evolution of the probe fields; additionally, the diffraction effect is negligible, which is valid for the probe fields having large transverse size.

It is easy to understand the basic feature of the propagation of the probe fields through solving the Maxwell-Bloch (MB) equations (\ref{BEs}) and (\ref{Atoms MEs}) with $\sigma_{l1}$ ($l=1,2,3$) and $\Omega_{pj}$ ($j=1,2$) proportional to the form $\exp [i(K_{a}z-\omega t)]$~\cite{note02}. We obtain
\begin{equation}
K_{a}^{\pm}\left(\omega\right)=\frac{\omega}{c}+\frac{-\left(\kappa_{14}D_{3}+\kappa_{13}D_{4}\right)\pm\sqrt{\left(\kappa_{14}D_{3}-\kappa_{13}D_{4}\right)^2+4\kappa_{13}\kappa_{14}|\Omega_{c1}\Omega_{c2}|^2}} {2\left[|\Omega_{c1}|^2\left(\omega+d_{41}\right)+|\Omega_{c2}|^2\left(\omega+d_{31}\right)-\left(\omega+d_{21}\right)\left(\omega+d_{31}\right)\left(\omega+d_{41}\right)\right]},
\label{Atoms DR}
\end{equation}
with $D_{3}=|\Omega_{c1}|^2-(\omega+d_{21})(\omega+d_{31})$ and $D_{4}=|\Omega_{c2}|^2-(\omega+d_{21})(\omega+d_{41})$.  We see that the MB equations allow two normal modes, with the linear dispersion relations given by $K_{a}^{+}$ and $K_{a}^{-}$, respectively.

Fig.~\ref{Fig1}(b) shows ${\rm Im}(K_{a}^{+})$ [i.e. the imaginary part of $K_{a}^{+}$] as a function of $\omega$ for $\Omega_{c1}=\Omega_{c2}=20\,{\rm MHz}$  (red dashed line) and $\Omega_{c1}=\Omega_{c2}=60\,{\rm MHz}$ (green dashed-dot line). When plotting the figure, $\Delta_{j}$ ($j=1,2,3$) are set to be zero, and realistic parameters from $^{87}{\rm Rb}$ atoms are taken, given by $\Gamma_{13}=\Gamma_{23}=\Gamma_{14}=\Gamma_{24}=16\,{\rm MHz}$, $\kappa_{13}=\kappa_{14}=1\times10^{10}\,{\rm cm}^{-3}$~\cite{Rb87}.
We see that a transparency window is opened in the profile of ${\rm Im}(K_{a}^{+})$ near $\omega=0$; the transparency window becomes larger when the control fields are increased. The opening of the transparency window (called EIT transparency window)
is due to the EIT effect contributed by the control fields. The blue solid cure in the figure is ${\rm Im}(K_{a}^{-})$ as a function of $\omega$, which however has always a large absorption peak near $\omega=0$ irrespective of the value of the control fields.
Below, for convenience we shall call the normal mode with the linear dispersion relation $K_{a}^{+}$  ($K_{a}^{-}$) as EIT-mode (non-EIT-mode).

The double-$\Lambda$-type four-level system can be used to describe a resonant FWM process in atomic systems~\cite{Fleischhauer,Merriam,DengPRA1,Kang,Ying}. The first laser field (i.e. the control field tuned to the $|2\rangle\leftrightarrow |3\rangle$ transition with the half Rabi frequency $\Omega_{c1}$) and the second laser field (i.e. the probe field tuned to the $|1\rangle \leftrightarrow |3\rangle$ transition with the half Rabi frequency $\Omega_{p1}$) can adiabatically establish a large atomic coherence of the Raman transition, described by the off-diagonal density matrix element $\sigma_{21}$. The third laser field, i.e. the control field tuned to the $|2\rangle\leftrightarrow |4\rangle$ transition with the half Rabi frequency $\Omega_{c2}$, can mix with the coherence $\sigma_{21}$ to generate a fourth field with the half Rabi frequency $\Omega_{p2}$ resonant with the $|1\rangle\leftrightarrow|4\rangle$ transition.
For details, see Refs.\cite{Fleischhauer,Merriam,DengPRA1,Kang,Ying} and references therein.

\subsection{Metamaterial analogue of the double-$\Lambda$-type four-level atomic system}

\begin{figure}
\includegraphics[width=0.9\columnwidth]{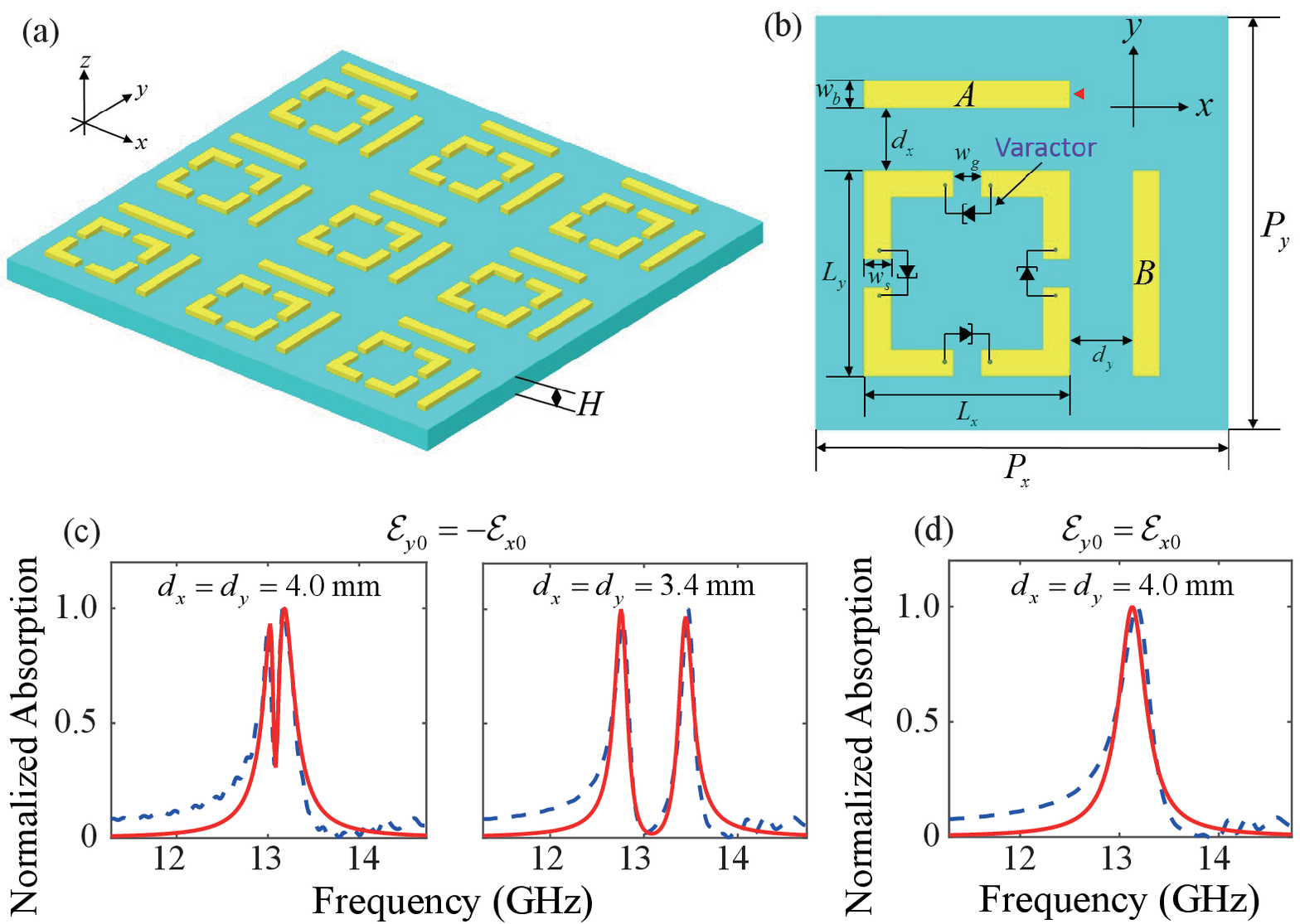}
\caption{(color online) (a) Schematic of the plasmonic metamaterial, which is an array of meta-atoms. (b) The meta-atom consists of two cut-wires (CWs) (indicated by ``A'' and ``B'') and a split-ring resonator (SRR), where the parameters $d_{x}$, $d_{y}$, $L_{x}$, $L_{y}$, $w_{b}$, $w_{g}$, and $w_{s}$ are given in the text. For generating nonlinear excitations, four hyperabrupt tuning varactors are mounted onto the slits of the SRR. (Sec.~\ref{sec3}).
(c) The numerical result (blue dashed lines)
of the normalized absorption spectrum of the EM wave as a function of frequency by taking ${\cal E}_{y0}=-{\cal E}_{x0}$,  $d_{x}=d_{y}=4.0\,{\rm mm}$ (first panel), and $d_{x}=d_{y}=3.4\,{\rm mm}$ (second panel). (d) The numerical result (blue dashed line) of normalized absorption spectrum for ${\cal E}_{y0}={\cal E}_{x0}$, $d_{x}=d_{y}=4.0\,{\rm mm}$.  Red
solid lines in (c) and (d) are analytical results obtained from the formula ${\rm Im}(q_{10})$ given by Eq.~(\ref{LEs Sols}) in Appendix~\ref{app2}. }
\label{Fig2}
\end{figure}

We now seek for a possible classical analogue of the above four-level atomic model by using a metamaterial, which
is assumed to be an array [Fig.~\ref{Fig2}(a)] of unit cells (i.e. meta-atoms) [Fig.~\ref{Fig2}(b)] consisting two CWs (indicated by ``A'' and ``B'') and a SRR.
The CW A and CW B are, respectively, positioned along $x$ and $y$ direction, while the SRR is formed by a square ring with a gap at the center of each side. 20-${\rm\mu m}$-thick copper forming the CWs and the SRR is etched on a substrate with a height of $H=1.5\,{\rm mm}$. Geometrical parameters of the meta-atom are $L_{x}=L_{y}=8\,{\rm mm}$, $w_{s}=1.2\,{\rm mm}$, $w_{g}=0.6\,{\rm mm}$, and $w_{b}=1.2\,{\rm mm}$~\cite{note1000}.

We assume that an incident gigahertz radiation ${\bf E}={\bf e}_x E_x+{\bf e}_y E_y$
[with $E_j={\cal E}_{j0}e^{-i\omega_p t}+{\rm c.c.}$ ($j=x,y$)] is collimated on the array of the meta-atoms, with  polarization component $E_x$  ($E_y$)  parallel to the CW A (CW B). In order to understand the EM property of the system, a numerical simulation is carried out by using the commercial finite difference time domain software package (CST Microwave Studio) and probing the imaginary part of the radiative field amplitude at the center of the end facet of the CW $A$ [red arrow in Fig.~\ref{Fig2}(b)]~\cite{Zhangs}. The blue dashed lines in Fig.~\ref{Fig2}(c) are
normalized absorption spectrum as a function of the incident wave frequency by taking the excitation condition
${\cal E}_{y0}=-{\cal E}_{x0}$ for $d_{x}=d_{y}=3.4\,{\rm mm}$ (first panel), and ${\cal E}_{y0}=-{\cal E}_{x0}$ for $d_{x}=d_{y}=4.0\,{\rm mm}$ (second panel). Here and below, ${\cal E}_{x0}$ and ${\cal E}_{y0}$ are taken to be real for simplicity. Fig.~\ref{Fig2}(d) shows the normalized absorption spectrum under the excitation condition  ${\cal E}_{y0}={\cal E}_{x0}$ for $d_{x}=d_{y}=4.0\,{\rm mm}$.
The red solid lines in the figure are analytical results obtained from ${\rm Im}(q_{10})$ based on Eq.~(\ref{LEs Sols}) in Appendix~\ref{app2}. We see that the absorption spectrum profile depends on excitation condition, which is quite different from the PIT absorption spectrum considered before.

The dependence on the excitation condition for the absorption spectrum can be briefly explained as follows. A sole CW in the meta-atoms is function as an optical dipole antenna and thus serves as a bright (or radiative) oscillator, which can be directly excited by the incident radiation. The surface current in an excited SRR can be clockwise or anticlockwise direction, indicating that there is no direct electric dipole coupling with the incident radiation and hence the SRR serves as a dark or trapped oscillator with long dephasing time~\cite{note03}. For the excitation condition ${\cal E}_{y0}=-{\cal E}_{x0}$  [Fig.~\ref{Fig2}(c)], the surface current is cooperatively induced through the near-field coupling between SRR and CWs, resulting in a maximum enhancement of the dark-oscillator resonance and thus the substantial suppression of the absorption of the incident radiation, acting like a typical PIT metamaterial. However, for the excitation condition ${\cal E}_{y0}={\cal E}_{x0}$ [Fig.~\ref{Fig2}(d)], the surface current is suppressed due to an opposite excitation direction, leading to a complete suppression of the dark-oscillator resonance. As a result, the radiation absorption is significant (acting like a sole CW) and hence no PIT behavior occurs.
For convenience, in the following we called the excitation mode under the ${\cal E}_{y0}=-{\cal E}_{x0}$ [Fig.~\ref{Fig2}(c)] as the PIT-mode, and the excitation mode under the ${\cal E}_{y0}={\cal E}_{x0}$ [Fig.~\ref{Fig2}(d)] as the non-PIT-mode (or absorption mode).

The dynamics of the bright oscillators (i.e. CW A and CW B) and dark oscillator (i.e. SRR) in the meta-atoms can be described by the coupled Lorentz equations~\cite{Zhangs,Gu,Chen,bai,Bai2,Bai3}
\begin{subequations}
\label{LEs}
\begin{align}
& \frac{\partial^2q_{1}}{\partial t^2}+\gamma_{1}\frac{\partial q_{1}}{\partial t}+\omega_{1}^2q_{1}-\kappa_{1}q_{3}=g_{1}E_{x},\\
& \frac{\partial^2q_{2}}{\partial t^2}+\gamma_{2}\frac{\partial q_{2}}{\partial t}+\omega_{2}^2q_{2}-\kappa_{2}q_{3}=g_{2}E_{y},\\
& \frac{\partial^2q_{3}}{\partial t^2}+\gamma_{3}\frac{\partial q_{3}}{\partial t}+\omega_{3}^2q_{3}-\kappa_{1}q_{1}-\kappa_{2}q_{2}=0,
\end{align}
\end{subequations}
where $q_{l}$ are displacements from the equilibrium position of the bright oscillators ($l=1,2$) and the dark oscillator ($l=3$), with $\gamma_{l}$  and $\omega_{l}$~\cite{note03} respectively the damping rate and the natural frequency of $l$th oscillator; $g_{1}$ ($g_{2}$) is the parameter describing the coupling between the CW A (CW B) and the $x$-polarization ($y$-polarization) component of the EM wave, and $\kappa_{1}$ ($\kappa_{2})$ is the  parameter describing the coupling between CW A (CW B) and SRR. 
The numerical values of these coefficients can be obtained from the numerical result presented in Fig.~\ref{Fig2}, by using the method described in Appendix~\ref{app2}.

Based on the solution given by Eq.~(\ref{LEs Sols}), we deduce that, in the case of $\omega_3=\omega_p$, $\gamma_{3}=0$ and $g_1=g_2$, Eq.~(\ref{LEs}) allows a ``dark state''  (i.e. the state where both the bright oscillators are not excited, i.e. $q_{10}=q_{20}=0$) exists, if
\begin{equation}\label{MM DS}
\kappa_{2}{\cal E}_{y0}-\kappa_{1}{\cal E}_{x0}=0.\
\end{equation}
This ``dark state'' condition is equivalent to the one obtained in the four-level double-$\Lambda$-type atomic system, given by Eq.~(\ref{Atoms Dark State}).
Obviously, the PIT-mode shown in Fig.~\ref{Fig2}(c) corresponds to the case $\kappa_2=-\kappa_1$, where the minus symbol can be understood as a $\pi$-phase difference resulting in a cooperative coupling effect, which is assumed in all the numerical calculations carried out below.

The equation of motion of the EM wave is governed by the Maxwell equation
\begin{equation}\label{MaxEqu}
\nabla^2 E_{x(y)}-\frac{1}{c^2}\frac{\partial^2 E_{x(y)}}{\partial t^2}=\frac{1}{\varepsilon_{0} c^2}\frac{\partial^2 P_{x(y)}}{\partial t^2},
\end{equation}
where $P_{x(y)}=\varepsilon_{0}\chi_{\rm D}^{(1)}E_{x(y)}+N_{m}eq_{1(2)}$, with $N_{m}$ the unit-cell density, $e$ the unit charge, and $\chi_{\rm D}^{(1)}$ the optical susceptibility of the hosting material.
We assume the distance between the meta-atoms is large so that the interaction between them can be neglected.

Assuming the central frequency of the incident radiation $\omega_{p}$ is near the natural frequencies of the Lorentz oscillators described by Eq.~(\ref{LEs})~\cite{note03}, a resonant interaction occurs between the incident radiation and these oscillators. To deal with the propagation problem of the plasmonic polaritons in the system analytically, we assume $E_{j}({\bf r},t)={\cal E}_{j}(z,t)e^{i\left(k_{p}z-\omega_{p}t\right)}+{\rm c.c.}$ and $q_{l}({\bf r},t)={\tilde q}_{l}(z,t)\exp[i(k_{p}z-\omega_{l}t-\Delta_{l}t)]+{\rm c.c.}$, where ${\cal E}_{j}(z,t)$ and ${\tilde q}_{l}(z,t)$ are slowly-varying envelopes and $\Delta_{l}=\omega_{p}-\omega_{l}$ is a small detuning. With
this ansatz and under RWA, Eq.~(\ref{LEs}) is simplified into the reduced Lorentz equation
\begin{subequations}\label{Reduced LEs}
\begin{align}
& \left(i\frac{\partial}{\partial t}+d_{1}\right){\tilde q}_{1}+\frac{\kappa_{1}}{2\omega_{p}}{\tilde q}_{3}+\frac{g_{1}}{2\omega_{p}}{\cal E}_{x}=0,\\
& \left(i\frac{\partial}{\partial t}+d_{2}\right){\tilde q}_{2}+\frac{\kappa_{2}}{2\omega_{p}}{\tilde q}_{3}+\frac{g_{2}}{2\omega_{p}}{\cal E}_{y}=0,\\
& \left(i\frac{\partial}{\partial t}+d_{3}\right){\tilde q}_{3}+\frac{\kappa_{1}}{2\omega_{p}}{\tilde q}_{1}+\frac{\kappa_{2}}{2\omega_{p}}{\tilde q}_{2}=0,
\end{align}
\end{subequations}
with $d_{j}=\Delta_{j}+\gamma_{j}/2$. We see that the reduced Lorentz equation (\ref{Reduced LEs}) describing the unit cell has the same form as the optical Bloch equation (\ref{BEs}) describing the four-level double-$\Lambda$ atom. Consequently, {\it each unit cell in the metamaterial is analogous to a four-level double-$\Lambda$-type atom in the atomic gas} presented in the last subsection. That is to say, the unit cell is indeed a meta-atom, where the bright-oscillator excitation in the CW A (CW B) driven by ${\cal E}_{x}$  (${\cal E}_{y}$) is equivalent to the dipole-allowed transition $|1\rangle\leftrightarrow|3\rangle$ ($|1\rangle\leftrightarrow|4\rangle$) driven by the probe field $\Omega_{p1}$ ($\Omega_{p2}$), and the dark-oscillator excitation in the SRR is equivalent to the dipole-forbidden transition $|1\rangle\rightarrow|2\rangle$ in the four-level double-$\Lambda$-type atom. We also see that
the coupling between the CW A (CW B) and the SRR, described by $\kappa_{1}$ ($\kappa_{2}$), is equivalent to the control field
$\Omega_{c1}$ ($\Omega_{c2}$) driven the atomic transition $|2\rangle\leftrightarrow|3\rangle$ ($|2\rangle\leftrightarrow|4\rangle$).

Under SVEA, the Maxwell equation in the metamaterial reads
\begin{subequations}\label{MMs MEs}
\begin{align}
i\left(\frac{\partial}{\partial z}+\frac{n_{\rm D}}{c}\frac{\partial}{\partial t}\right){\cal E}_{x}+\kappa_{0}{\tilde q}_{1}=0,\\
i\left(\frac{\partial}{\partial z}+\frac{n_{\rm D}}{c}\frac{\partial}{\partial t}\right){\cal E}_{y}+\kappa_{0}{\tilde q}_{2}=0,
\end{align}
\end{subequations}
with $\kappa_{0}=N_{m}e\omega_{p}/(2\varepsilon_{0}cn_{\rm D})$, $n_{\rm D}=\sqrt{1+\chi_{\rm D}^{(1)}}$.  Obviously,
Eq.~(\ref{MMs MEs}) has similar structure as Eq.~(\ref{Atoms MEs}). Thus, a complete correspondence between the MB equations (\ref{BEs}) and (\ref{Atoms MEs}) described the four-level double-$\Lambda$-type atomic gas and the Maxwell-Lorentz (ML) equations (\ref{Reduced LEs}) and (\ref{MMs MEs}) described the plasmonic metamaterial is established.

The propagation feature of a plasmonic polariton in the metamaterial can be obtained by
assuming all quantities in the ML equations (\ref{Reduced LEs}) and (\ref{MMs MEs}) proportional to $\exp[i(K_{m}z-\omega t)]$. It is easy to get the linear dispersion relation
\begin{equation}
K_{m}^{\pm}\left(\omega\right)=\frac{n_{\rm D}}{c}\omega+\kappa_{0}\frac{-\left(R_{1}g_{f2}+R_{2}g_{f1}\right)\pm\sqrt{\left(R_{1}g_{f2}-R_{2}g_{f1}\right)^2
+4\kappa_{f1}^2\kappa_{f2}^2 g_{f1}g_{f2}}} {2\left[\kappa_{f1}^2\left(\omega+d_{2}\right)+\kappa_{f2}^2\left(\omega+d_{1}\right)-\left(\omega+d_{3}\right)\left(\omega+d_{1}\right)\left(\omega+d_{2}\right)\right]},
\label{MMs DR}
\end{equation}
where $R_{j}=\kappa_{fj}^2-(\omega+d_{j})(\omega+d_{3})$, with $\kappa_{fj}=\kappa_{j}/(2\omega_{p})$ and $g_{fj}=g_{j}/(2\omega_{p})$  ($j=1,2$). As expected, the metamaterial system allows two normal modes with the linear dispersion relation respectively given by $K_{m}^{+}$ and $K_{m}^{-}$. In fact, $K_{m}^{+}$ ($K_{m}^{-}$) is a PIT-mode (non-PIT-mode) of the system, as explained below.

\begin{figure}
\includegraphics[width=0.85\columnwidth]{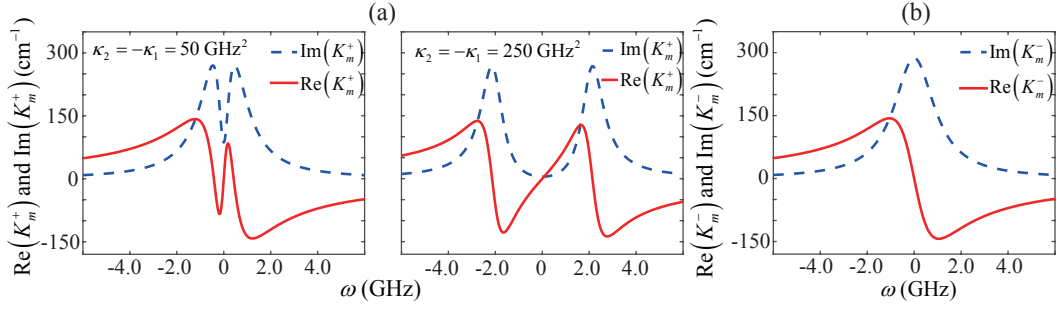}
\caption{(color online) (a) Linear dispersion relation of the $K_{m}^{+}$-mode (PIT-mode). Im($K_{m}^{+}$) (blue dashed line) and Re($K_{m}^{+}$) (red solid line) are plotted as functions of $\omega$ for $\kappa_{2}=-\kappa_{1}=50\,{\rm GHz}^2$ (first panel) and $\kappa_{2}=-\kappa_{1}=250\,{\rm GHz}^2$ (second panel). (b) Linear dispersion relation of the $K_{m}^{-}$-mode (non-PIT-mode) for arbitrary $\kappa_{1}$ ($\kappa_{2}=-\kappa_{1}$). }
\label{Fig3}
\end{figure}

The character of the above two normal modes can be clearly illustrated by plotting $K_{m}^{+}$ and $K_{m}^{-}$ as functions of $\omega$. Shown in Fig.~\ref{Fig3}(a)
are Im($K_{m}^{+}$) (blue dashed line) and Re($K_{m}^{+}$) (red solid line) for $\kappa_{2}=-\kappa_{1}=50\,{\rm GHz}^2$ (first panel; corresponding to $d_{x}=d_{y}=4.0\,{\rm mm}$) and $\kappa_{2}=-\kappa_{1}=250\,{\rm GHz}^2$ (second panel; corresponding to $d_{x}=d_{y}=3.4\,{\rm mm}$), respectively.
When plotting the figure, the system parameters are taken from Appendix~\ref{app2}, and additional parameters are chosen by $\kappa_{0}=10^{10}\,{\rm kg/(cm\cdot s^{2}\cdot C)}$~\cite{Mikhail} and $\Delta_{j}=0\,(j=1,2,3)$.
We see that ${\rm Im}(K_{m}^{+})$ displays a transparency window (called PIT transparency window) near $\omega=0$, analogous to the EIT transparency window in ${\rm Im}(K_{a}^{+})$ of the four-level double-$\Lambda$-type atomic system [red dashed line and green dashed-dot line in Fig.~\ref{Fig1}(b)]. The steep slope of ${\rm Re}(K_{m}^{+})$ indicates a normal dispersion and a slow group velocity of the plasmonic polariton. As the coupling strength between the CWs and the SRR gets larger (i.e. the separations $d_{x}$ and $d_{y}$ is reduced), the PIT transparency window becomes wider and deeper, and the slope of ${\rm Re}(K_{m}^{+})$ gets flatter. The opening of the PIT transparency window is attributed to the destructive interference between the two bright oscillators and the dark oscillator through cooperative near-field coupling. Shown in Fig.~\ref{Fig3}(b) is the imaginary (red solid line) and the
real (blue dashed line) of $K_{m}^{-}$, which is nearly independent on the coupling constant $\kappa_{1}$  ($\kappa_{2}=-\kappa_{1}$). We see that ${\rm Im}(K_{m}^{-})$ has a single, large absorption peak and ${\rm Re}(K_{m}^{-})$ has an abnormal dispersion near $\omega=0$, analogous to ${\rm Im}(K_{a}^{-})$ of the double-$\Lambda$-type atomic system [blue solid line in Fig.~\ref{Fig1}(b)].

\subsection{Propagation of linear plasmonic polaritons via an analogous FWM process of atomic system}\label{Sec.IIC}

As indicated above, the meta-atoms in the present metamaterial system are analogous to the four-level atoms with the  double-$\Lambda$-type configuration, and hence an analogous resonant FWM phenomenon for the plasmonic polaritons is possible. That is to say, if initially only one polarization-component of the EM wave (e.g. $x$-component) is injected into the metamaterail, a new polarization-component (e.g. $y$-component) will be generated through two equivalent control fields (i.e. the couplings between the SRR and CWs, described by $\kappa_1$ and $\kappa_2$). To illustrate this, we present the solution of the ML equations (\ref{Reduced LEs}) and (\ref{MMs MEs})
\begin{subequations}
\begin{align}
& {\cal E}_{x}(z,t)=\frac{1}{2\pi}\int_{-\infty}^{+\infty}d\omega\left[F_{0}^{+}e^{i(K_{m}^{+}z-\omega t)}+F_{0}^{-}e^{i(K_{m}^{-}z-\omega t)}\right],\\
& {\cal E}_{y}(z,t)=\frac{1}{2\pi}\int_{-\infty}^{+\infty}d\omega\left[G^{+}F_{0}^{+}e^{i(K_{m}^{+}z-\omega t)}+G^{-}F_{0}^{-}e^{i(K_{m}^{-}z-\omega t)}\right],
\end{align}
\label{Linear Sols}
\end{subequations}
which can be obtained by using Fourier transform~\cite{Ying,LiHJ}. Here $G^{\pm}=[-A\pm(A^2+4\kappa_{f1}^2\kappa_{f2}^2g_{f1}g_{f2})^{1/2}]/(2\kappa_{f1}\kappa_{f2}g_{f2})$ with $A=R_{1}g_{f2}-R_{2}g_{f1}$, and $F_{0}^{\pm}$ is the initial amplitude of the normal mode $K_{m}^{\pm}$ determined by given excitation condition. We assume initially only the $x$-component of the EM field in input to the system, i.e.
the initial condition for the EM field is given by ${\cal E}_{x}(0,t)\neq 0$, ${\cal E}_{y}(0,t)=0$. By Eq.~(\ref{Linear Sols}) we have
\begin{subequations}\label{FWM}
\begin{align}
& {\cal E}_{x}(z,t)=\frac{1}{2\pi}\int_{-\infty}^{+\infty}d\omega\frac{G^{+}e^{i(K_{m}^{-}z-\omega t)}-G^{-}e^{i(K_{m}^{+}z-\omega t)}}{G^{+}-G^{-}}{\cal\tilde E}_{x}(0,\omega),\\
& {\cal E}_{y}(z,t)=\frac{1}{2\pi}\int_{-\infty}^{+\infty}d\omega \frac{G^{+}G^{-}}{G^{+}-G^{-}}\left[e^{i(K_{m}^{-}z-\omega t)}-e^{i(K_{m}^{+}z-\omega t)}\right]{\cal\tilde E}_{x}(0,\omega),
\end{align}
\end{subequations}
where ${\cal\tilde E}_{x}(0,\omega)=\int^{+\infty}_{-\infty}dt{\cal E}_{x}(0,t)e^{i\omega t}$. For simplicity,
we consider the adiabatic regime where the power series of $K_{m}^{\pm}$ and $G^{\pm}$ on $\omega$ converge rapidly. By taking $K_{m}^{\pm}=K_{0}^{\pm}+\omega/V_{g}^{\pm}+O(\omega^2)$ and $G^{\pm}=G_{0}^{\pm}+O(\omega)$, we readily obtain
\begin{subequations}\label{FWM Sols}
\begin{align}
& {\cal E}_{x}(z,t)=\frac{G_{0}^{+}{\cal E}_{x}(0,\tau_{-})e^{iK_{0}^{-}z}-G_{0}^{-}{\cal E}_{x}(0,\tau_{+})e^{iK_{0}^{+}z}}{G_{0}^{+}-G_{0}^{-}},\\
& {\cal E}_{y}(z,t)=\frac{G_{0}^{+}G_{0}^{-}}{G_{0}^{+}-G_{0}^{-}}\left[{\cal E}_{x}(0,\tau_{-})e^{iK_{0}^{-}z}-{\cal E}_{x}(0,\tau_{+})e^{iK_{0}^{+}z}\right],
\end{align}
\end{subequations}
where $\tau_{\pm}=t-z/V_{g}^{\pm}$, with $V_{g}^{\pm}\equiv (\partial K_{m}^{\pm}/\partial\omega)^{-1}|_{\omega=0}$ being the group-velocity of the normal mode $K_{m}^{\pm}$.

\begin{figure}
\includegraphics[width=0.48\columnwidth]{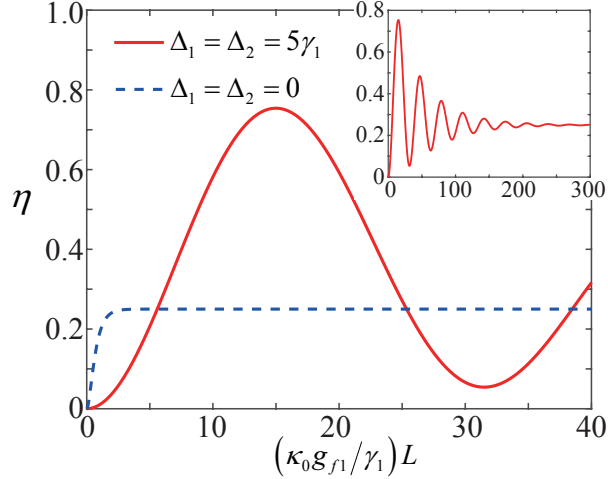}
\caption{(color online) FWM conversion efficiency $\eta$ as a function of the dimensionless optical depth $(\kappa_{0}g_{f1}/\gamma_{1})L$ for $\Delta_{1}=\Delta_{2}=0$ (blue dashed line), and $\Delta_{1}=\Delta_{2}=5\gamma_{1}$ (red solid line). Inset: FWM conversion efficiency $\eta$ for optical depth up to 300 for $\Delta_{1}=\Delta_{2}=5\gamma_{1}$. }
\label{Fig4}
\end{figure}

The conversion efficiency of the FWM is given by $\eta(L)\equiv\int_{-\infty}^{+\infty}dt|{\cal E}_{y}(L,t)/{\cal E}_{x}(0,t)|^2$, where $L$ is the medium length. For the case $\kappa_{2}=-\kappa_{1}$, one has ${\rm Im}(K_{0}^{-})\gg{\rm Im}(K_{0}^{+})$, which means that the $K_{m}^{-}$ mode decays away rapidly during propagation and hence can be safely neglected. Then Eq.~(\ref{FWM Sols}) is simplified as
\begin{subequations}\label{FWM}
\begin{align}
& {\cal E}_{x}(z,t)=\frac{G_{0}^{-}}{G_{0}^{-}-G_{0}^{+}}{\cal E}_{x}(0,\tau_{+})e^{iK_{0}^{+}z},\\
& {\cal E}_{y}(z,t)=\frac{G_{0}^{-}G_{0}^{+}}{G_{0}^{-}-G_{0}^{+}}{\cal E}_{x}(0,\tau_{+})e^{iK_{0}^{+}z}.
\end{align}
\end{subequations}
We see that the $x$-and $y$-polarization components of the EM wave have matched group velocity $V_{g}^{+}$. The expression of the FWM conversion efficiency reduces into
\begin{equation}
\eta(L)=\frac{|G_{0}^{+}G_{0}^{-}|^2}{|G_{0}^{+}-G_{0}^{-}|^2}|\exp(iK_{0}^{+}L)|^2.
\end{equation}
Shown in Fig.~\ref{Fig4}
is the FWM conversion efficiency $\eta$ as a function of the dimensionless optical depth $(\kappa_{0}g_{f1}/\gamma_{1})L$ for $\Delta_{1}=\Delta_{2}=0$ (blue dashed line) and for $\Delta_{1}=\Delta_{2}=5\gamma_{1}$ (red solid line). When plotting this figure, we have set $\Delta_{3}=0$ and $\gamma_{3}\approx 0$ in order for a better analogue to the atomic system. The influence of $\gamma_{3}$ can be effectively reduced by introducing a gain element into the gaps of the SRRs (see the discussion in Sec.~\ref{sec4}). From the figure, we see that for the case of exact resonance (i.e. $\Delta_{1}=\Delta_{2}=0$), the FWM efficiency $\eta$ increases and rapidly saturates to $25\%$ when the dimensionless optical depth $(\kappa_{0}g_{f1}/\gamma_{1})L\approx 5$ (i.e. $L\approx 0.9\,{\rm cm}$), indicating a unidirectional energy transmission from ${\cal E}_{x}$ to ${\cal E}_{y}$. For the case of far-off resonance (i.e. $\Delta_{1}=\Delta_{2}=5\gamma_{1}$), the FWM efficiency displays a damped oscillation in the interval $0<(\kappa_{0}g_{f1}/\gamma_{1})L<250$, indicating a back-and-forth energy exchange between ${\cal E}_{x}$ and ${\cal E}_{y}$; eventually the efficiency reach to the steady-state value $25\%$ when $(\kappa_{0}g_{f1}/\gamma_{1})L\ge 300$ (see the inset). Interestingly, the value of the FWM conversion efficiency may reach to $\eta\approx76\%$ at $(\kappa_{0}g_{f1}/\gamma_{1})L \approx 15$ (i.e. $L\approx3\,{\rm cm}$).
%

%

\section{Vector plasmonic dromions in the PIT metamaterial}\label{sec3}

Note that when deriving Eq.~(\ref{MMs MEs}), the diffraction effect has been neglected, which is invalid for the plasmonic polaritons with small transverse size or long propagation distance; furthermore, because of the highly resonant (and hence dispersive) character inherent in the PIT metamaterial, the linear plasmonic polaritons obtained above inevitably undergo significant distortion during propagation. Hence it is necessary to seek the possibility to obtain a robust propagation of the plasmonic polaritons in the PIT metamaterial. One way to solve this problem is to make the PIT system work in a nonlinear propagation regime.

In recent years, nonlinear metamaterials have attracted much attention due to their potential applications (see Ref.~\cite{Lapine} and references therein). One suitable way to design a nonlinear PIT metamaterial in microwave and lower THz ranges is to use nonlinear insertions onto the meta-atoms~\cite{Shadrivov,Wang}.
Here, as suggested in Refs.~\cite{Shadrivov,Wang,Bai2}, we assume the nonlinear insertion in the PIT metamaterial are varactor diodes, which are mounted onto the gaps of the SRRs~\cite{Wang} [see Fig.~\ref{Fig2}(b)].

\subsection{Nonlinear envelope equations}
Since the introduction of the nonlinear element onto the SRRs, Eq.~(\ref{LEs}c) should be replaced by~\cite{Wang}
\begin{equation}\label{Non LEs}
\frac{\partial^2q_{3}}{\partial t^2}+\gamma_{3}\frac{\partial q_{3}}{\partial t}+\omega_{3}^2q_{3}-\kappa_{2}q_{1}-\kappa_{2}q_{2}+\alpha q_{3}^2+\beta q_{3}^3=0,
\end{equation}
where $\alpha$ and $\beta$ are nonlinearity coefficients, described in Appendix~\ref{app5}.

Due to the quadratic and cubic nonlinearities in Eq.~(\ref{Non LEs}), the input EM field (with only a fundamental wave) will generate longwave (rectification),
and second harmonic components, i.e. $E_{l}({\bf r},t)={\cal E}_{dl}({\bf r},t)+[{\cal E}_{fl}({\bf r},t)e^{i(k_{p}z-\omega_{p}t)}+{\rm c.c.}]+[{\cal E}_{sl}({\bf r},t)e^{i\theta_p}+{\rm c.c.}]$ ($l=x,y$), with $\theta_p=(2k_{p}+\Delta k)z-2\omega_{p}t$ and $\Delta k$ a detuning in wavenumber. The oscillations of the Lorentz oscillators in the meta-atoms have the form
$q_{j}({\bf r},t)=q_{dj}({\bf r},t)+[q_{fj}({\bf r},t)e^{i\theta_j}+{\rm c.c.}]+[q_{sj}({\bf r},t)e^{2i\theta_j}+{\rm c.c.}]$ ($j=1,2,3$), with $\theta_j=k_{j}z-\omega_{j}t-\Delta_{j}t$.
Substituting these expressions into Eqs.~(\ref{LEs}a), (\ref{LEs}b), (\ref{MaxEqu}), and (\ref{Non LEs}), and adopting RWA and SVEA, we obtain a series of equations for the motion of $q_{\mu j}$ and ${\cal E}_{\mu l}$, listed in Appendix~\ref{app3}.

We solve the equations for $q_{\mu j}$ and $E_{\mu l}$ by using the standard method of multiple scales~\cite{Jeffery}.
Take the asymptotic expansion $q_{fj}=\epsilon q_{fj}^{(1)}+\epsilon^2q_{fj}^{(2)}+\cdots$, $q_{dj}=\epsilon^2q_{dj}^{(2)}+\cdots$, $q_{sj}=\epsilon^2q_{sj}^{(2)}+\cdots$, ${\cal E}_{fl}=\epsilon {\cal E}_{fl}^{(1)}+\epsilon^2 {\cal E}_{fl}^{(2)}+\cdots$, and
${\cal E}_{dl}=\epsilon^2 {\cal E}_{dl}^{(2)}+\cdots$ (here $\epsilon$ is a dimensionless small parameter characterizing the amplitude of the incident EM field), and assume all quantities on the right sides of the asymptotic expansion as functions of the multiscale variables~\cite{Jeffery} $z_{l}=\epsilon^{l}z\,(l=0,1,2)$ and $t_{l}=\epsilon^{l}t\,(l=0,1)$. Substituting the expansion into the equations for $q_{\mu j}$ and ${\cal E}_{\mu j}$ and comparing powers of $\epsilon$, we obtain a chain of linear but inhomogeneous equations which can be solved order by order.

The leading order [i.e. $O(\epsilon)$] solution reads ${\cal E}_{fx}^{(1)}=F_{+}e^{i\theta_{+}}$ and ${\cal E}_{fy}^{(1)}=G^{+}F_{+}e^{i\theta_{+}}$, where $\theta_{+}=K_{m}^{+}z_{0}-\omega t_{0}$ and $F_{+}$ is a slowly-varying envelope function to be determined in higher-order approximations. The expression of
$G^{+}$ is given in Sec.~\ref{Sec.IIC}. Here we consider only the PIT (i.e. $K_{m}^{+}$) mode because the non-PIT (i.e. $K_{m}^{-}$) mode decays rapidly during propagation, as indicated in the last section. The solution for $q_{\mu j}^{(1)}$  is presented in Appendix~\ref{app4}.

At the second order [i.e. $O(\epsilon^2)$], a solvability condition yields $i[\partial/\partial z_{1}+(1/V_{g}^{+})\partial/\partial t_{1}]F_{+}=0$, where $V_{g}^{+}$ is the group-velocity of the fundamental wave. 
Solution for the longwave is ${\cal E}_{dx}^{(2)}=Q_{+}$ and ${\cal E}_{dy}^{(2)}=G^{+}Q_{+}$, with $Q_{+}$  the slowly-varying envelope to be determined yet. Explicit expressions of the solutions for other quantities at this order are listed in Appendix~\ref{app4}.

At the third order [i.e. $O(\epsilon^3)$], a solvability condition results in the equation
\begin{equation}
i\frac{\partial F_{+}}{\partial z_{2}}-\frac{K_{2}^{+}}{2}\frac{\partial^2F_{+}}{\partial t_{1}^2}+\frac{c}{2\omega_{p}n_{\rm D}}\left(\frac{\partial^2}{\partial x_{1}^2}+\frac{\partial^2}{\partial y_{1}^2}\right)F_{+}+\frac{\omega_{p}R_{0}}{2cn_{\rm D}}\chi_{+}^{(2)}Q_{+}F_{+}+\frac{\omega_{p}}{2cn_{\rm D}}\chi_{+}^{(3)}\left|F_{+}\right|^2F_{+}=0,
\label{Third Order}
\end{equation}
where $K_{2}^{+}\equiv[\partial^2 K_{m}^{+}/\partial\omega^2]|_{\omega=0}$ describes the group-velocity dispersion of the fundamental wave; $R_{0}$ is a coefficient characterizing the coupling between the fundamental and long waves,
with the expression given in Appendix~\ref{app4}; $\chi_{+}^{(2)}$ and $\chi_{+}^{(3)}$ are, respectively, the second-order and third-order nonlinear susceptibilities of the $K_{m}^{+}$ mode, with the form
\begin{subequations}\label{chi}
\begin{align}
\chi_{+}^{(2)}=&\frac{N_{m}e}{\varepsilon_{0}}\frac{2\alpha(\omega_{2}^2\kappa_{1}g_{1}+\omega_{1}^2\kappa_{2}g_{2}G^{+})}{g_{2}G^{+}(G^{+}-G^{-})(\omega_{1}^2\omega_{2}^2\omega_{3}^2-\omega_{2}^2\kappa_{1}^2-\omega_{1}^2\kappa_{2}^2)}
\left|\frac{D_{2}\kappa_{1}g_{1}+D_{1}\kappa_{2}g_{2}G^{+}}{D_{1}D_{2}D_{3}-D_{2}\kappa_{1}^2-D_{1}\kappa_{2}^2}\right|^2,\\
\nonumber \chi_{+}^{(3)}=&\frac{N_{m}e}{\varepsilon_{0}}\frac{\left(D_{2}\kappa_{1}g_{1}+D_{1}\kappa_{2}g_{2}G^{+}\right)^2\left|D_{2}\kappa_{1}g_{1}+D_{1}\kappa_{2}g_{2}G^{+}\right|^2} {g_{2}G^{+}(G^{+}-G^{-})(D_{1}D_{2}D_{3}-D_{2}\kappa_{1}^2-D_{1}\kappa_{2}^2)^2\left|D_{1}D_{2}D_{3}-D_{2}\kappa_{1}^2-D_{1}\kappa_{2}^2\right|^2}\\
& \times\left[\left(\frac{4\alpha^2\omega_{1}^2\omega_{2}^2}{\omega_{1}^2\omega_{2}^2\omega_{3}^2-\omega_{2}^2\kappa_{1}^2
-\omega_{1}^2\kappa_{2}^2}+\frac{2\alpha^2H_{1}H_{2}}{H_{1}H_{2}H_{3}-H_{2}\kappa_{1}^2-H_{1}\kappa_{2}^2}\right)
-3\beta\right].
\end{align}
\end{subequations}

At the fourth order [i.e. $O(\epsilon^4)$], a solvability condition results in the equation for the longwave $Q_{+}$
\begin{equation}\label{Forth Order}
\left(\frac{\partial^2}{\partial x_{1}^2}+\frac{\partial^2}{\partial y_{1}^2}\right)Q_{+}+\left[\left(\frac{1}{V_{g}^{+}}\right)^2-\left(\frac{1}{V_{p}^{+}}\right)^2\right]\frac{\partial^2Q_{+}}{\partial t_{1}^2}-\frac{\chi_{+}^{(2)}}{c^2}\frac{\partial^2|F_{+}|^2}{\partial t_{1}^2}=0,
\end{equation}
here $V_{p}^{+}\equiv [n_{m}^{+}(0)/c]^{-1}$ is the phase-velocity of the longwave, defined by
$n_{m}^{+}(0)=n_{m}^{+}|_{\omega_{p}=0,\omega=0}$ with $n_{m}^{+}(\omega;\omega_{p})=c[k_{p}(\omega_{p})+K_{m}^{+}(\omega;\omega_{p})]/(\omega_{p}+\omega)$.
It is easy to obtain
\begin{equation}
\frac{1}{V_{p}^{+}}=\frac{n_{\rm D}}{c}+\frac{N_{m}e}{2\varepsilon_{0}cn_{\rm D}}\frac{(g_{1}X_{2}+g_{2}X_{1})+\sqrt{(g_{1}X_{2}-g_{2}X_{1})^2+4\kappa_{1}^2\kappa_{2}^2g_{1}g_{2}}}{\omega_{1}^2\omega_{2}^2\omega_{3}^2-\omega_{1}^2\kappa_{2}^2-\omega_{2}^2\kappa_{1}^2},
\end{equation}
with $X_{j}=\omega_{j}\omega_{3}-\kappa_{j}^2$.

Under the PIT condition [i.e. $(\kappa_{j}/2\omega_{p})^2\gg\gamma_{j}\gamma_{3}/4$, $j=1,2$], the real parts of the nonlinearity susceptibilities $\chi_{+}^{(2)}$ and $\chi_{+}^{(3)}$ are greatly enhanced and their imaginary parts are much smaller than their real parts. Interestingly, $\chi_{+}^{(3)}$ can be further enhanced via a strong coupling between the longwave and the fundamental wave. This can be seen from the expression of the effective third-order nonlinearity susceptibility of the $K_{m}^{+}$ mode~\cite{Newell}
\begin{equation}
\chi_{\rm eff}^{(3)}=\chi_{+}^{(3)}+\frac{R_{0}\left[\chi_{+}^{(2)}\right]^2}{c^2\left[\left(\frac{1}{V_{p}^{+}}\right)-\left(\frac{1}{V_{g}^{+}}\right)^2\right]},
\label{chieff}
\end{equation}
obtained by neglecting the diffraction term in Eq.~(\ref{Forth Order}) and then plugging the derived expression for $Q_{+}$ into Eq.~(\ref{Third Order}). The second term in Eq.~(\ref{chieff}) is due to the {\it longwave-shortwave interaction}, where one can clearly see that when the group-velocity of the shortwave is nearly equal to the phase-velocity of the longwaves (i.e. $V_{g}^{+}\approx V_{p}^{+}$), the effective third-order nonlinear susceptibility $\chi_{\rm eff}^{(3)}$ can be further enhanced, which can be realized by adjusting the coupling parameters $\kappa_{1}$ and $\kappa_{2}$. Compared with the conventional PIT-based metamaterials~\cite{bai,Bai2} with one bright and one dark oscillators, the present FWM-based metamaterial consisting double bright and one dark oscillators allows for a stronger bright-dark coupling, resulting in an enhanced Kerr nonlinearity. As a result, the coupling strength for the longwave-shortwave interaction in the present FWM-based metamaterial is $\sqrt{2}$ time larger than that in the conventional PIT-based metamaterials.
%

\subsection{Vector plasmonic dromions}

Equations (\ref{Third Order}) and~(\ref{Forth Order}) show that the self-interaction of the shortwave $F_{+}$ can stimulate the generation of the longwave $Q_{+}$ [Eq.~(\ref{Forth Order})], and at the same time the longwave  $Q_{+}$ has a back-action to the shortwave $F_{+}$ [Eq.~(\ref{Third Order})]. Such equations admit solutions describing the excitation of high-dimensional, two-component (vector) nonlinear plasmonic polaritons in the PIT metamaterial. To demonstrate this, we convert them into the dimensionless form
\begin{subequations}\label{Dimensionless}
\begin{align}
& i\frac{\partial u}{\partial s}+\left(\frac{\partial^2}{\partial\xi^2}+g_{d0}\frac{\partial^2}{\partial\eta^2}+g_{d1}\frac{\partial^2}{\partial\tau^2}\right)u+2g_{1}|u|^2u+g_{2}vu=0,\\
& g_{d2}\frac{\partial^2v}{\partial\tau^2}-\left(\frac{\partial^2}{\partial\xi^2}+g_{d0}\frac{\partial^2}{\partial\eta^2}\right)v+g_{3}\frac{\partial^2|u|^2}{\partial\tau^2}=0.
\end{align}
\end{subequations}
where $u=\epsilon F_{+}/U_{0}$, $v=\epsilon^2Q_{+}/V_{0}$, $s=z/(2L_{\rm diff})$, $\tau=(t-z/V_{g}^{+})/\tau_{0}$, $\xi=x/R_{x}$, $\eta=y/R_{y}$, and $g_{d0}=(R_{x}/R_{y})^2$, $g_{d1}=L_{\rm diff}/L_{\rm disp}$, $g_{d2}=[R_{x}^2/\tau_{0}^2][(1/V_{p}^{+})^2-(1/V_{g}^{+})^2]$, $g_{1}=L_{\rm diff}/L_{\rm nln}$, $g_{2}=L_{\rm diff}[\omega_{p}/(n_{\rm D}c)]R_{0}V_{0}\chi_{+}^{(2)}$, $g_{3}=R_{x}^2U_{0}^2\chi_{+}^{(2)}/[c^2\tau_{0}^2V_{0}]$. Here $R_{x}\,(R_{y})$ is the typical radius of the incident EM field in the $x\,(y)$ direction; $\tau_{0}$ is the typical pulse duration of the probe field; $U_{0}\,(V_{0})$ is the typical amplitude of the longwave (shortwave) envelope; $L_{\rm disp}=-\tau_{0}^2/{\rm Re}(K_{2}^{+})$, $L_{\rm diff}\equiv n_{\rm D}\omega_{p}R_{x}^2/c$ and $L_{\rm nln}=(2n_{\rm D}c)/[\omega_{p}U_{0}^2{\rm Re}(\chi_{+}^{(3)})]$ are, respectively, the typical dispersion length, the typical diffraction length, and the typical nonlinearity length. Note that in obtaining Eq.~(\ref{Dimensionless}), we have neglected the small imaginary parts of $\chi_{+}^{(3)}$ and $K_{2}^{+}$, which is reasonable under the PIT condition as discussed above.

In favor of the formation of plasmonic dromions, we take the following two assumptions. First, we shall assume $R_{x}\ll R_{y}$ and thus $g_{d0}\ll1$ so that the original (3+1)-dimensional nonlinear problem can be reduced into a (2+1)-dimensional one. Second, we assume the contribution of the dispersion, diffraction and the nonlinearity effects are of the same level, which can be achieved by taking $L_{\rm diff}=L_{\rm disp}=L_{\rm nln}$ and thus we obtain $\tau_{0}=R_{x}\sqrt{-\omega_{p}n_{\rm D}{\rm Re}(K_{2}^{+})/c}$ and $U_{0}=[c/(\omega_{p}R_{x})]\sqrt{2/{\rm Re}[\chi_{+}^{(3)}]}$. In fact, these two assumptions can be realized by taking the realistic set of parameters, i.e. $\Delta_{1}=\Delta_{2}=-5\gamma_{1}$, $\kappa_{2}=-\kappa_{1}=4260\,{\rm GHz}^{2}$, $R_{x}=1.8\,{\rm cm}$, $R_{y}=10.2\,{\rm cm}$, $U_{0}=8.25\,{\rm V/cm}$, $V_{0}=1.3\,{\rm V/cm}$, $\tau_{0}=4.1\times10^{-11}\,{\rm s}$, and hence we have $K_{2}^{+}=(-1.30+i0.10)\times10^{-19}\,{\rm cm^{-2}s}$, $\chi_{+}^{(3)}=1.20\times10^{-3}+i2.26\times10^{-7}\,{\rm cm^2/V^2}$, $L_{\rm diff}=L_{\rm disp}=L_{\rm nln}=13.4\,{\rm cm}$, $g_{1}=g_{2}=g_{d1}=g_{d2}=1$, $g_{3}=4$, and $g_{d0}\ll1$. For such case, Eq.~(\ref{Dimensionless}) can be simplified into standard Davey-Stewartson-I (DS-I) equation
\begin{subequations}\label{DSE_dimensionless}
\begin{align}
& i\frac{\partial u}{\partial s}+\frac{\partial^2u}{\partial\xi_{1}^2}+\frac{\partial^2u}{\partial\tau_{1}^2}+v_{1}u=0,\\
& \left(\frac{\partial^2}{\partial\xi_{1}^2}+\frac{\partial^2}{\partial\tau_{1}^2}\right)|u|^2=\frac{\partial^2v_{1}}{\partial\xi_{1}\partial\tau_{1}},
\end{align}
\end{subequations}
with $v_{1}=v+2|u|^2$, where, for convenience, we have performed a 45-degree rotation of coordinates $\xi_{1}=(\xi+\tau)/\sqrt{2}$ and $\tau_{1}=(\xi-\tau)/\sqrt{2}$. The DS-I equation (\ref{DSE_dimensionless}) can be exactly solved via the Hirota's bilinear method~\cite{Hirota} and various dromion solutions can be obtained. A single-dromion solution reads~\cite{Dromion} $u=G/F$ and $v_{1}=V_{11}+V_{12}$ with $V_{11}=2\partial^2{\rm ln}(F)/\partial\xi_{1}^2$ and $V_{12}=2\partial^2{\rm ln}(F)/\partial\tau_{1}^2$, and
\begin{subequations}\label{Dromion Sols}
\begin{align}
F&=1+\exp(\eta_{1}+\eta_{1}^{\ast})+\exp(\eta_{2}+\eta_{2}^{\ast})+\gamma\exp(\eta_{1}+\eta_{1}^{\ast}+\eta_{2}+\eta_{2}^{\ast}),\\
G&=\rho\exp(\eta_{1}+\eta_{2}),
\end{align}
\end{subequations}
where $\eta_{j}=(k_{jr}+ik_{ji})(r_{j}-r_{j0})+(\Omega_{jr}+i\Omega_{ji})s\,(j=1,2)$, with $(r_{1(0)},r_{2(0)})=(\xi_{1(0)},\tau_{1(0)})$, $\Omega_{jr}=-2k_{jr}k_{ji}$, $\Omega_{1i}+\Omega_{2i}=k_{1r}^2+k_{2r}^2-k_{1i}^2-k_{2i}^2$, and $\rho=2\sqrt{2k_{1r}k_{2r}(\gamma-1)}$. Here $k_{jr}$, $k_{ji}$, $r_{j0}$ and $\gamma$ are free real parameters.

\begin{figure}
\includegraphics[width=0.75\columnwidth]{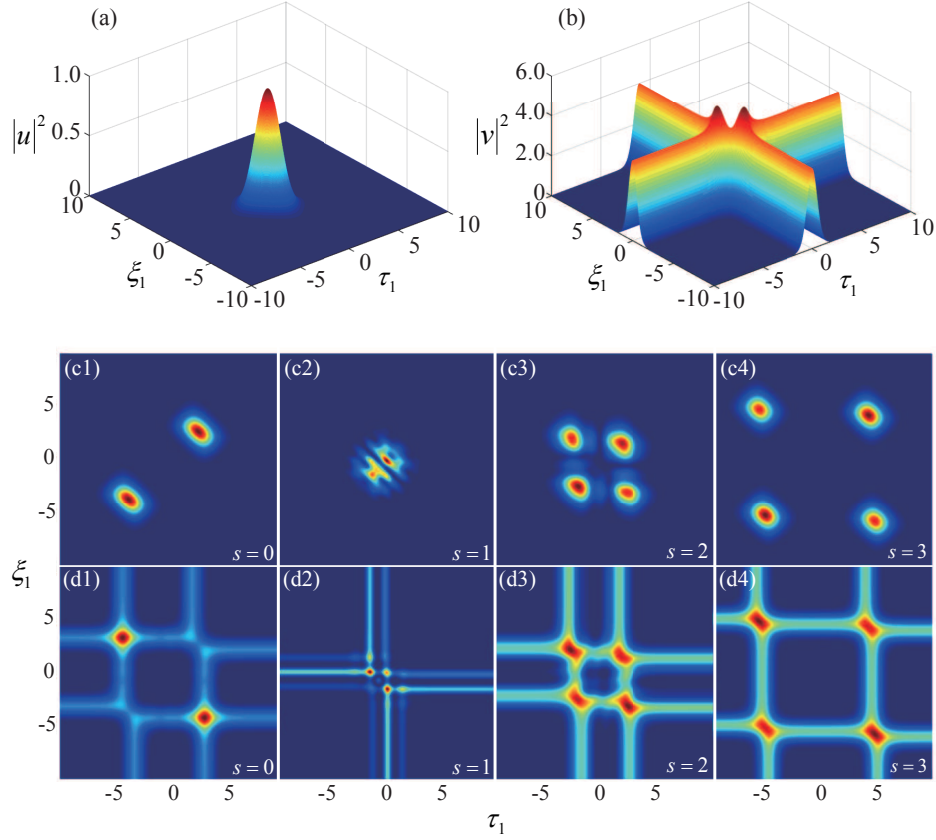}
\caption{(color online) Plasmonic dromions and their interaction. (a) [(b)] is the intensity profile of the shortwave $|u|^2$ (longwave $|v_1|^2$) as functions of $\xi_{1}$ and $\tau_{1}$ at $s=0$. (c1), (c2), (c3), (c4) [(d1), (d2), (d3), (d4)] are intensity profiles of the shortwave $|u|^2$ (longwave $|v_1|^2$) during the interaction between two dromions, respectively at $s\equiv z/(2L_{\rm diff})=0,1,2,3$. System parameters are given in the text.}
\label{Fig5}
\end{figure}

Shown in Fig.~\ref{Fig5}(a) and Fig.~\ref{Fig5}(b) are, respectively, intensity distributions of the shortwave profile $|u|^2$ and the longwave profile $|v_1|^2$ as functions of $\xi_{1}$ and $\tau_{1}$ at $s=0$, when taking $\gamma=9$, $\xi_{10}=\tau_{10}=0$, $k_{1r}=k_{2r}=1$, and $k_{1i}=k_{2i}=0$. Obviously, the shortwave profile $|u|^2$ denotes a localized envelope function, which decays exponentially in all spatial directions [Fig.~\ref{Fig5}(a)]; the longwave profile $|v_1|^2$ denotes two interacting plane solitons (kinks), which decay in their respective traveling directions [Fig.~\ref{Fig5}(b)]. Such high-dimensional
nonlinear excitation is called dromion~\cite{Hirota}.

We proceed with the investigation on the interaction between two plasmonic dromions. In doing so, we first integrate Eq.~(\ref{DSE_dimensionless}b), yielding
\begin{equation}\label{Potential}
v_{1}(\tau_{1},\xi_{1},s)=\int_{-\infty}^{\tau_{1}}\frac{\partial|u|^2}{\partial \xi_{1} }d\tau_{1}+\int_{-\infty}^{\xi_{1}}\frac{\partial|u|^2}{\partial \tau_{1}}d\xi_{1}+V_{11}|_{\tau_{1}\rightarrow -\infty}+V_{12}|_{\xi_{1}\rightarrow -\infty},
\end{equation}
where the nonzero boundary conditions $V_{11}|_{\tau_{1}\rightarrow -\infty}$ and $V_{12}|_{\xi_{1}\rightarrow -\infty}$ are given by~\cite{Dromion}
\begin{subequations}\label{BC}
\begin{align}
& V_{11}|_{\tau_{1}\rightarrow -\infty}=2k_{1r}^2{\rm sech}^2[k_{1r}(\xi_{1}-\xi_{10})+\Omega_{1r}s], \\
& V_{12}|_{\xi_{1}\rightarrow -\infty}=2k_{2r}^2{\rm sech}^2[k_{2r}(\tau_{1}-\tau_{10})+\Omega_{2r}s].
\end{align}
\end{subequations}
In the numerical simulation, the integral Eq.~(\ref{Potential}) is conducted by fourth-order Runge-Kutta method with the boundary condition~(\ref{BC}), and then Eq.~(\ref{DSE_dimensionless}a) with the obtained result $v_{1}$ is solved by using  split-step Fourier method. Shown in Fig.~\ref{Fig5}(c1)-Fig.~\ref{Fig5}(c4) [Fig.~\ref{Fig5}(d1)-Fig.~\ref{Fig5}(d4)] are numerical results for the evolution of the shortwave $|u|^2$ (longwave $|v_1|^2$) during the collision between two dromions at  $s\equiv z/(2L_{\rm diff})=0,1,2,3$, respectively.  When doing the simulation we have taken a superposition of two dromion solutions as an initial input [i.e. Fig.~\ref{Fig5}(c1) and Fig.~\ref{Fig5}(d1)], and initial speeds and positions of the two dromions are setting to be $k_{1i}=-k_{2i}=1.8$ and $r_{10}=-r_{20}=3.2$. From the figure we see that two initial dromions become four dromions after the collision, which are, respectively, located around the four intersections of the longwave $v_{1}$, and these four dromions gradually separate and propagate almost stably, indicating that the collision between dromions is inelastic. The reason is that the four intersections of $v_{1}$ are the most attractive points in the entire region, which attract the EM wave intensities of the main peaks of the shortwaves $u$ during the collision, resulting in the appearance of four pulses for the shortwave located around the four intersections after the collision~\cite{Dromion}.

Within the forth-order approximation, the explicit expression for the EM field in the metamaterial takes the form
\begin{equation}\label{vectordromion}
{\bf E}\left({\bf r},t\right) \equiv \frac{{\bf e}_{x}\kappa_{1}+{\bf e}_{y}\kappa_{2}}{\kappa_{1}^2+\kappa_{2}^2}\left[\left(U_{0}ue^{ik_{p}z-i\omega_{p} t}+{\rm c.c.}\right)+V_{0}v\right].
\end{equation}
%
When $u$ and $v$ are taken as the dromion solution given above, we obtain a vector plasmonic dromion since the EM field (\ref{vectordromion}) has two polarization components, with each component a plasmonic dromion.
Note that, different from the result in the scalar model considered before~\cite{Bai2}, the polarization of the EM field obtained here can be actively selected by adjusting the separation between the CWs and SRR [i.e. $d_{x}$ and $d_{y}$ in Fig.~\ref{Fig2}(b) and hence the coupling constants $\kappa_{1}$ and $\kappa_{2}$], which can be served as a polarization selector for practical applications~\cite{Pol_Sel1,Pol_Sel2}.

The threshold of the power density of the vector plasmonic dromion given above can be estimated by using Poynting vector. Based on the above system parameters, the average power of the vector plasmonic dromion is estimated as
\begin{equation}
{\bar P}=6.1\,{\rm mW}.
\end{equation}
We see that due to the resonant character of the PIT effect in the system, extremely low generation power is required for generating the vector plasmonic dromion.

\section{Discussion and summary}\label{sec4}

It should be mentioned that in writing the dark-state condition (\ref{MM DS}), the damping coefficient $\gamma_{3}$ of the dark oscillator in the meta-atoms is assumed to be small. However, due to the Ohmic loss inherent in the metal that construct the metamaterial, by our numerical calculation the numerical value of $\gamma_{3}$ is about $0.18\,{\rm GHz}$, which, though smaller than damping coefficients of the CWs ($\gamma_{1}=\gamma_{2}=2.1\,{\rm GHz}$), is still large and has inevitably detrimental impact on the PIT quality. In order to improve the performance of the PIT, one can suppress $\gamma_{3}$ by introducing a gain element into the SRR of the meta-atoms. One possible way is the use of tunneling diodes that have negative resistance and hence may provide gain to the PIT-based metamaterial~\cite{Jiang,Ye}. Such method has been recognized to be useful for suppressing and even cancelling $\gamma_{3}$, particularly in microwave and THz regimes.

In conclusion, in this article we have considered a plasmonic metamaterial interacting with an EM field with two polarization components. We have proved that such metamaterial can be taken as a classical analogue of an atomic gas with a double-$\Lambda$-type four-level configuration coupled with four laser fields, displays an PIT effect and an equivalent process of atomic FWM. We have shown that, when the nonlinear varactors are mounted onto the gaps of the SRRs, the metamaterial system can acquire giant second- and third-order Kerr nonlinearities via the PIT and the longwave-shortwave interaction. We have also shown that the system supports high-dimensional vector plasmonic dromions, which have very low generation power and are robust during propagation. Our work not only contributes a plasmonic analogue of atomic EIT and FWM but also provides a way for generating novel plasmonic polaritons, and hence opens a new avenue on the exploration of PIT effect in metamaterials.

\section*{Acknowledgments}

The authors thank Shuang Zhang for helpful discussions. This work was supported by the NSF-China under Grant No.~11474099.

\appendix
\section{Explicit Expressions of the atomic Bloch equation}\label{app1}
Explicit expressions of the Bloch equation for the density-matrix elements $\sigma_{jl}$ of the four-level double-$\Lambda$-type atoms are given by
\begin{subequations}\label{ExMB1}
\begin{align}
& i\frac{\partial\sigma_{11}}{\partial t}-i\Gamma_{13}\sigma_{33}-i\Gamma_{14}\sigma_{44}-\Omega_{p1}\sigma_{31}^{\ast}-\Omega_{p2}\sigma_{41}^{\ast}+\Omega_{p1}^{\ast}\sigma_{31}
+\Omega_{p2}^{\ast}\sigma_{41}=0,\\
& i\frac{\partial\sigma_{22}}{\partial t}-i\Gamma_{23}\sigma_{33}-i\Gamma_{24}\sigma_{44}-\Omega_{c1}\sigma_{32}^{\ast}-\Omega_{c2}\sigma_{42}^{\ast}+\Omega_{c1}^{\ast}\sigma_{32}
+\Omega_{c2}^{\ast}\sigma_{42}=0,\\
& i\left(\frac{\partial}{\partial t}+\Gamma_{13}+\Gamma_{23}\right)\sigma_{33}+\Omega_{p1}\sigma_{31}^{\ast}+\Omega_{c1}\sigma_{32}^{\ast}-\Omega_{p1}^{\ast}\sigma_{31}-\Omega_{c1}^{\ast}\sigma_{32}=0,\\
& i\left(\frac{\partial}{\partial t}+\Gamma_{14}+\Gamma_{24}\right)\sigma_{44}+\Omega_{p2}\sigma_{41}^{\ast}+\Omega_{c2}\sigma_{42}^{\ast}
-\Omega_{p2}^{\ast}\sigma_{41}-\Omega_{c2}^{\ast}\sigma_{42}=0
\end{align}
\end{subequations}
for diagonal elements, and
\begin{subequations}\label{ExMB2}
\begin{align}
& \left(i\frac{\partial}{\partial t}+d_{21}\right)\sigma_{21}-\Omega_{p1}\sigma_{32}^{\ast}-\Omega_{p2}\sigma_{42}^{\ast}+\Omega_{c1}^{\ast}\sigma_{31}+\Omega_{c2}^{\ast}\sigma_{41}=0,\\
& \left(i\frac{\partial}{\partial t}+d_{31}\right)\sigma_{31}-\Omega_{p1}\left(\sigma_{33}-\sigma_{11}\right)-\Omega_{p2}\sigma_{43}^{\ast}+\Omega_{c1}\sigma_{21}=0,\\
& \left(i\frac{\partial}{\partial t}+d_{32}\right)\sigma_{32}-\Omega_{p2}\left(\sigma_{44}-\sigma_{11}\right)-\Omega_{c2}\sigma_{43}^{\ast}+\Omega_{p1}\sigma_{21}^{\ast}=0,\\
& \left(i\frac{\partial}{\partial t}+d_{41}\right)\sigma_{41}-\Omega_{c1}\left(\sigma_{33}-\sigma_{22}\right)-\Omega_{p1}\sigma_{43}+\Omega_{c2}\sigma_{21}=0,\\
& \left(i\frac{\partial}{\partial t}+d_{42}\right)\sigma_{42}-\Omega_{c2}\left(\sigma_{44}-\sigma_{22}\right)-\Omega_{c1}\sigma_{43}+\Omega_{p2}\sigma_{21}^{\ast}=0,\\
& \left(i\frac{\partial}{\partial t}+d_{43}\right)\sigma_{43}-\Omega_{p1}^{\ast}\sigma_{41}-\Omega_{c1}^{\ast}\sigma_{42}+\Omega_{p2}\sigma_{31}^{\ast}+\Omega_{c2}\sigma_{32}^{\ast}=0,
\end{align}
\end{subequations}
for non-diagonal elements, where $\Gamma_{j}=\sum_{l<j}\Gamma_{jl}$, with $\Gamma_{jl}$ the spontaneous emission decay rate from the state $|l\rangle$ to the state $|j\rangle$; $d_{jl}=\Delta_{j}-\Delta_{l}+i\gamma_{jl}$, $\gamma_{jl}=(\Gamma_{j}+\Gamma_{l})/2+\gamma_{jl}^{\rm dep}$, with $\gamma_{jl}^{\rm dep}$ the dephasing rate between the state $|j\rangle$ and the state $|l\rangle$.

\section{Determination of the system parameters in the Lorentz equation}\label{app2}
The coefficients in the coupled Lorentz equation Eq.~(\ref{LEs}) is determined by fitting the numerical result in
Fig.~\ref{Fig2}(c) and Fig.~\ref{Fig2}(d) obtained by using the finite difference time domain software package (CST Microwave Studio) stated in the main text, and the analytical result of Eq.~(\ref{LEs}). Assuming the solution of Eq.~(\ref{LEs}) has the form $q_{l}(z,t)=q_{l0}\exp(ik_{p}z-i\omega_{p} t)+{\rm c.c.}\,(l=1,2,3)$ and $E_{j}(z,t)={\cal E}_{j0}\exp(ik_{p}z-i\omega_{p} t)+{\rm c.c.}\,(j=x,y)$, we have
\begin{subequations}\label{LEs Sols}
\begin{align}
q_{10}&=\frac{\left(D_{3}D_{2}-\kappa_{2}^2\right)g_{1}{\cal E}_{x0}+\kappa_{1}\kappa_{2}g_{2}{\cal E}_{y0}} {D_{1}D_{2}D_{3}-D_{2}\kappa_{1}^2-D_{1}\kappa_{2}^2},\\
q_{20}&=\frac{\kappa_{2}\kappa_{1}g_{1}{\cal E}_{x0}+\left(D_{3}D_{1}-\kappa_{1}^2\right)g_{2}{\cal E}_{y0}} {D_{1}D_{2}D_{3}-D_{2}\kappa_{1}^2-D_{1}\kappa_{2}^2},\\
q_{30}&=\frac{D_{2}\kappa_{1}g_{1}{\cal E}_{x0}+D_{1}\kappa_{2}g_{2}{\cal E}_{y0}}{D_{1}D_{2}D_{3}-D_{2}\kappa_{1}^2-D_{1}\kappa_{2}^2},
\end{align}
\end{subequations}
with $D_{j}=\omega_{j}^2-\omega_{p}^2-i\gamma_{j}\omega_{p}$.

The red solid lines in Fig.~\ref{Fig2}(c) and Fig.~\ref{Fig2}(d) (where ${\cal E}_{x0}$  and ${\cal E}_{y0}$ have been taken to be real) show the analytical result based on Eq.~(\ref{LEs Sols}a) in the cases ${\cal E}_{y0}=-{\cal E}_{x0}$ (the excitation condition of the PIT-mode) and ${\cal E}_{y0}={\cal E}_{x0}$ (the excitation condition of the non-PIT-mode), respectively. A better fitting yields $\omega_{1}=\omega_{2}=2\pi\times 13.13\,{\rm GHz}$, $\omega_{3}=2\pi\times 13.08\,{\rm GHz}$
$\gamma_{1}=\gamma_{2}=2.1\,{\rm GHz}$, $\gamma_{3}=0.15\,{\rm GHz}$, $\kappa_{2}=-\kappa_{1}=50\,{\rm GHz}^{2}$ for $d_{x}=d_{y}=4.0\,{\rm mm}$ and $\kappa_{2}=-\kappa_{1}=250\,{\rm GHz}^{2}$ for $d_{x}=d_{y}=3.4\,{\rm mm}$, and $g_{1}=g_{2}=1.79\times10^{11}\,{\rm C/kg}$.

\section{Equations for $q_{\mu j}$ and ${\cal E}_{\mu j}$}\label{app3}
The equations of motion for $q_{\mu j}$ read
\begin{subequations}\allowdisplaybreaks\label{Reduced Non LEs}
\begin{eqnarray}
&& \left(i\frac{\partial}{\partial t}+d_{f1}\right)q_{f1}+\frac{\kappa_{1}}{2\omega_{p}}q_{f3}+\frac{g_{1}}{2\omega_{p}}{\cal E}_{fx}=0,\\
&& \left(i\frac{\partial}{\partial t}+d_{f2}\right)q_{f2}+\frac{\kappa_{2}}{2\omega_{p}}q_{f3}+\frac{g_{2}}{2\omega_{p}}{\cal E}_{fy}=0,\\
&& \left(i\frac{\partial}{\partial t}+d_{f3}\right)q_{f3}+\frac{\kappa_{1}}{2\omega_{p}}q_{f1}+\frac{\kappa_{2}}{2\omega_{p}}q_{f2}\nonumber \\
&& \vspace{2cm} -\frac{1}{2\omega_{p}}\left[2\alpha(q_{d3}q_{f3}+q_{s3}q_{f3}^{\ast})+3\beta|q_{f3}|^2q_{f3}\right]=0,\\
&& \left(\frac{\partial^2}{\partial t^2}+\gamma_{1}\frac{\partial}{\partial t}+\omega_{1}^2\right)q_{d1}-\kappa_{1}q_{d3}-g_{1}{\cal E}_{dx}=0,\\
&& \left(\frac{\partial^2}{\partial t^2}+\gamma_{2}\frac{\partial}{\partial t}+\omega_{2}^2\right)q_{d2}-\kappa_{2}q_{d3}-g_{2}{\cal E}_{dy}=0,\\
&& \left(\frac{\partial^2}{\partial t^2}+\gamma_{3}\frac{\partial}{\partial t}+\omega_{3}^2\right)q_{d3}-\kappa_{2}q_{d1}-\kappa_{2}q_{d2}+2\alpha\left|q_{f3}\right|^2=0,\\
&& \left(i\frac{\partial}{\partial t}+d_{s1}+\frac{3}{4}\omega_{1}\right)q_{s1}+\frac{\kappa_{1}}{4\omega_{p}}q_{s3}+\frac{g_{1}}{4\omega_{p}}{\cal E}_{sx}e^{i\Delta kz}=0,\\
&& \left(i\frac{\partial}{\partial t}+d_{s2}+\frac{3}{4}\omega_{2}\right)q_{s2}+\frac{\kappa_{2}}{4\omega_{p}}q_{s3}+\frac{g_{2}}{4\omega_{p}}{\cal E}_{sy}e^{i\Delta kz}=0,\\
&& \left(i\frac{\partial}{\partial t}+d_{s3}+\frac{3}{4}\omega_{3}\right)q_{s3}+\frac{\kappa_{1}}{4\omega_{p}}q_{s1}+\frac{\kappa_{2}}{4\omega_{p}}q_{s2}-\frac{\alpha}{4\omega_{p}} q_{f3}^2=0,
\end{eqnarray}
\end{subequations}
with $d_{fj}=\Delta_{j}+i\gamma_{j}/2$ and $d_{sj}=2\Delta_{j}+i\gamma_{j}/2$, and equations for ${\cal E}_{\mu j}$ are given by
\begin{subequations}\allowdisplaybreaks
\begin{eqnarray}
&& i\left(\frac{\partial}{\partial z}+\frac{n_{\rm D}}{c}\frac{\partial}{\partial t}\right){\cal E}_{fx}+\frac{c}{2\omega_{p}n_{\rm D}}\left(\frac{\partial^2}{\partial x^2}+\frac{\partial^2}{\partial y^2}\right){\cal E}_{fx}+\kappa_{0}q_{f1}=0,\\
&& i\left(\frac{\partial}{\partial z}+\frac{n_{\rm D}}{c}\frac{\partial}{\partial t}\right){\cal E}_{fy}+\frac{c}{2\omega_{p}n_{\rm D}}\left(\frac{\partial^2}{\partial x^2}+\frac{\partial^2}{\partial y^2}\right){\cal E}_{fy}+\kappa_{0}q_{f2}=0,\\
&& \left(\frac{\partial^2}{\partial z^2}-\frac{n_{\rm D}^2}{c^2}\frac{\partial^2}{\partial t^2}\right){\cal E}_{dx}+\left(\frac{\partial^2}{\partial x^2}+\frac{\partial^2}{\partial y^2}\right){\cal E}_{dx}-\frac{N_{m}e}{\varepsilon_{0}c^2}\frac{\partial^2}{\partial t^2}q_{d1}=0,\\
&& \left(\frac{\partial^2}{\partial z^2}-\frac{n_{\rm D}^2}{c^2}\frac{\partial^2}{\partial t^2}\right){\cal E}_{dy}+\left(\frac{\partial^2}{\partial x^2}+\frac{\partial^2}{\partial y^2}\right){\cal E}_{dy}-\frac{N_{m}e}{\varepsilon_{0}c^2}\frac{\partial^2}{\partial t^2}q_{d2}=0,\\
&& i\left(\frac{\partial}{\partial z}+\frac{n_{\rm D}}{c}\frac{\partial}{\partial t}\right){\cal E}_{sx}+\frac{c}{4\omega_{p}n_{\rm D}}\left(\frac{\partial^2}{\partial x^2}+\frac{\partial^2}{\partial y^2}\right){\cal E}_{sx}+2\kappa_{0}q_{s1}e^{-i\Delta kz}=0,\\
&& i\left(\frac{\partial}{\partial z}+\frac{n_{\rm D}}{c}\frac{\partial}{\partial t}\right){\cal E}_{sy}+\frac{c}{4\omega_{p}n_{\rm D}}\left(\frac{\partial^2}{\partial x^2}+\frac{\partial^2}{\partial y^2}\right){\cal E}_{sy}+2\kappa_{0}q_{s2}e^{-i\Delta kz}=0.
\end{eqnarray}
\end{subequations}

\section{Solutions of $q_{\mu j}^{(m)} $ at each order approximations}\label{app4}
The first-order [i.e. $O(\epsilon)$] solution has only the fundamental wave, which reads
\begin{subequations}
\begin{align}
& q_{f1}^{(1)}=\frac{1}{\kappa_{0}}\left(K_{m}^{+}-\frac{n_{\rm D}}{c}\omega\right)F_{+}e^{i\theta_{+}},\\
& q_{f2}^{(1)}=\frac{G^{+}}{\kappa_{0}}\left(K_{m}^{+}-\frac{n_{\rm D}}{c}\omega\right)F_{+}e^{i\theta_{+}},\\
&
q_{f3}^{(1)}=\frac{D_{2}\kappa_{1}g_{1}+D_{1}\kappa_{2}g_{2}G^{+}}{D_{1}D_{2}D_{3}-D_{2}\kappa_{1}^2-D_{1}\kappa_{2}^2}F_{+}e^{i\theta_{+}},
\end{align}
\label{First Order Sols}
\end{subequations}
with $D_{j}=-2\omega_{j}(\omega+d_{fj})\,(j=1,2,3)$.

At the second order [i.e. $O(\epsilon^2)$], solution for the fundamental wave is given by
\begin{subequations}
\begin{align}
q_{f1}^{(2)}&=\frac{1}{\kappa_{0}}\left(\frac{\partial K_{m}^{+}}{\partial\omega}-\frac{n_{\rm D}}{c}\right)\left(i\frac{\partial}{\partial t_{1}}\right) F_{+}e^{i\theta_{+}},\\
q_{f2}^{(2)}&=\frac{G^{+}}{\kappa_{0}}\left(\frac{\partial K_{m}^{+}}{\partial\omega}-\frac{n_{\rm D}}{c}\right)\left(i\frac{\partial}{\partial t_{1}}\right) F_{+}e^{i\theta_{+}}.
\end{align}
\end{subequations}
The expression of $q_{f3}^{(2)}$ is long and omitted for saving space. Solution for the longwave (rectification) reads
\begin{subequations}
\begin{align}
q_{d1}^{(2)}&=\left(\frac{N_{m}e}{2\varepsilon_{0}n_{\rm D}}\right)^{-1}\left[n_{m}^{+}(0)-n_{\rm D}\right]Q_{+}+\frac{-2\alpha\omega_{2}^2\kappa_{1}}{\omega_{1}^2\omega_{2}^2\omega_{3}^2-\omega_{2}^2\kappa_{1}^2-\omega_{1}^2\kappa_{2}^2}\left|q_{f3}^{(1)}\right|^2,\\
q_{d2}^{(2)}&=\left(\frac{N_{m}e}{2\varepsilon_{0}n_{\rm D}}\right)^{-1}G^{+}\left[n_{m}^{+}(0)-n_{\rm D}\right]Q_{+}+\frac{-2\alpha\omega_{1}^2\kappa_{2}}{\omega_{1}^2\omega_{2}^2\omega_{3}^2-\omega_{2}^2\kappa_{1}^2-\omega_{1}^2\kappa_{2}^2}\left|q_{f3}^{(1)}\right|^2,\\
q_{d3}^{(2)}&=\frac{\omega_{2}^2\kappa_{1}g_{1}+\omega_{1}^2\kappa_{2}g_{2}G^{+}}{\omega_{1}^2\omega_{2}^2\omega_{3}^2-\omega_{2}^2\kappa_{1}^2-\omega_{1}^2\kappa_{2}^2}Q_{+}+\frac{-2\alpha\omega_{1}^2\omega_{2}^2\left|q_{f3}^{(1)}\right|^2}{\omega_{1}^2\omega_{2}^2\omega_{3}^2-\omega_{2}^2\kappa_{1}^2-\omega_{1}^2\kappa_{2}^2},
\end{align}
\end{subequations}
where $n_{m}^{+}(0)=n_{m}^{+}|_{\omega_{p}=0,\omega=0}$, with $n_{m}^{+}(\omega;\omega_{p})=c[k_{p}(\omega_{p})+K_{m}^{+}(\omega;\omega_{p})]/(\omega_{p}+\omega)$ the linear refractive index of the metamaterial. Solutions of the second-harmonic wave is
\begin{equation}
q_{s3}^{(2)}=\frac{-\alpha H_{1}H_{2}}{H_{1}H_{2}H_{3}-H_{2}\kappa_{1}^2-H_{1}\kappa_{2}^2}[q_{f3}^{(1)}]^2,
\end{equation}
with $H_{j}=-4\omega_{j}(2\omega+d_{sj}+3\omega_{j}/4)$, where $\delta_{ji}$ is the Kronecker symbol. Expressions for $q_{s1}^{(2)}$ and $q_{s2}^{(2)}$ are omitted here for saving space.

At the third order [i.e. $O(\epsilon^3)$], solution of the fundamental wave reads
\begin{subequations}
\begin{align}
q_{f1}^{(3)}&=\frac{-1}{2\kappa_{0}}\frac{\partial^2 K_{m}^{+}}{\partial\omega^2}\frac{\partial^2F_{+}}{\partial t_{1}^2}e^{i\theta_{+}}+\frac{D_{2}\kappa_{1}\left[2\alpha[q_{d3}^{(2)}q_{f3}^{(1)}+q_{s3}^{(2)}q_{f3}^{(1)\ast}]
+3\beta|q_{f3}^{(1)}|^2q_{f3}^{(1)}\right]}{D_{1}D_{2}D_{3}-D_{2}\kappa_{1}^2-D_{1}\kappa_{2}^2},\\
q_{f2}^{(3)}&=\frac{-G^{+}}{2\kappa_{0}}\frac{\partial^2 K_{m}^{+}}{\partial\omega^2}\frac{\partial^2F_{+}}{\partial t_{1}^2}e^{i\theta_{+}}+\frac{D_{1}\kappa_{2}\left[2\alpha[q_{d3}^{(2)}q_{f3}^{(1)}+q_{s3}^{(2)}q_{f3}^{(1)\ast}]
+3\beta|q_{f3}^{(1)}|^2q_{f3}^{(1)}\right]}{D_{1}D_{2}D_{3}-D_{2}\kappa_{1}^2-D_{1}\kappa_{2}^2}.
\end{align}
\end{subequations}
%
%

The expression of the coefficient $R_{0}$ in Eq.~(\ref{Third Order}) is given by
\begin{equation}
R_{0}=\frac{\left(D_{2}\kappa_{1}g_{1}+D_{1}\kappa_{2}g_{2}G^{+}\right)^2}{\left|D_{2}\kappa_{1}g_{1}+D_{1}\kappa_{2}g_{2}G^{+}\right|^2}
\frac{\left|D_{1}D_{2}D_{3}-D_{2}\kappa_{1}^2-D_{1}\kappa_{2}^2\right|^2}{\left(D_{1}D_{2}D_{3}-D_{2}\kappa_{1}^2-D_{1}\kappa_{2}^2\right)^2}.
\label{R0}
\end{equation}

\section{Nonlinear coefficients in Eq.~(\ref{Non LEs})}\label{app5}
The nonlinear property of the SRRs has been theoretically analyzed and experimentally measured in Ref.~\cite{Wang}. The value of $q$ in Ref.~\cite{Wang} represents the renormalized voltage, which has the unit of volt (V), while the value of $q_{3}$ in our work represents the amplitudes of the dark modes, which has the unit of centimeters (cm).
To make a comparison we switch the unit of Eq.~(\ref{LEs}), reading
\begin{subequations}
\begin{align}
& \frac{\partial^2u_{1}}{\partial t^2}+\gamma_{1}\frac{\partial u_{1}}{\partial t}+\omega_{1}^2u_{1}-\kappa_{1}u_{3}=\frac{g_{1}}{Q_{0}}E_{x},\\
& \frac{\partial^2u_{2}}{\partial t^2}+\gamma_{2}\frac{\partial u_{2}}{\partial t}+\omega_{2}^2u_{2}-\kappa_{2}u_{3}=\frac{g_{2}}{Q_{0}}E_{y},\\
& \frac{\partial^2u_{3}}{\partial t^2}+\gamma_{3}\frac{\partial u_{3}}{\partial t}+\omega_{3}^2u_{3}-\kappa_{1}u_{1}-\kappa_{2}u_{2}+\alpha Q_{0}u_{3}^2+\beta Q_{0}^2u_{3}^3=0,
\end{align}
\label{LEs1}
\end{subequations}
where $q_{j}=Q_{0}u_{j}\,(j=1,\,2,\,3)$. $u_{j}$ has the unit of V and $Q_{0}$ has the unit of $\rm cm\cdot V^{-1}$. Thus, nonlinear coefficients $\alpha$ and $\beta$ in our model~(\ref{LEs}) can be calculated through the dimension transformation $\alpha=Q_{0}^{-1}\alpha_{0}$ and $\beta=Q_{0}^{-2}\beta_{0}$, where $\alpha_{0}=-M\omega_{3}^2/(2V_{p})$ and $\beta_{0}=M(2M-1)\omega_{3}^2/(6V_{p}^2)$ can be found in Ref.~\cite{Wang}, readily given by $\alpha_{0}=-2.3503\times10^{3}\,{\rm V^{-1}GHz^{2}}$ and $\beta_{0}=4.8211\times10^{2}\,{\rm V^{-2}GHz^{2}}$ with $\omega_{3}=2\pi\times13.08\,{\rm GHz}$ given in context. Taking a typical value $Q_{0}=1.0\times10^{-9}\,{\rm V^{-1}cm}$~\cite{bai,Bai2} for dimension transformation, we finally obtain $\alpha=-2.3503\times10^{12}\,{\rm cm^{-1}GHz^{2}}$ and $\beta=4.8211\times10^{20}\,{\rm cm^{-2}GHz^{2}}$.
%


%


\begin{references}
\bibitem{Fleischhauer} M. Fleischhauer, A. Imamoglu, and J. P. Marangos,
Electromagnetically induced transparency: Optics in coherent media,
Rev. Mod. Phys. {\bf 77}, 633 (2005).

%
\bibitem{Alzar} C. L. G. Alzar, M. A. G. Martinez, and P. Nussenzveig,
Classical analog of electromagnetically induced transparency,
Am. J. Phys. {\bf 70}, 37 (2002).

\bibitem{Spring} A. G. Litvak and M. D. Tokman,
Electromagnetically Induced Transparency in Ensembles of Classical Oscillators,
Phys. Rev. Lett. {\bf 88}, 095003 (2002).

\bibitem{Harden}J. Harden, A. Joshi, and J. D. Serna,
Demonstration of double EIT using coupled harmonic oscillators and RLC circuits,
Eur. J. Phys. {\bf 32}, 541 (2011).

\bibitem{Souza}J. A. Souza, L. Cabral, R. R. Oliveira and C. J. Villas-Boas,
Electromagnetically-induced-transparency-related phenomena and their mechanical analogs,
Phys. Rev. A {\bf 92}, 023818 (2015).

\bibitem{Weis}S. Weis, R. Rivi${\rm \grave{e}}$re, S. Del${\rm \grave{e}}$glise, E. Gavartin, O. Arcizet, A. Schliesser, and T. J. Kippenberg,
Optomechanically induced transparency,
Science {\bf 330}, 1520 (2010).

\bibitem{Kronwald}A. Kronwald and F. Marquardt,
Optomechanically Induced Transparency in the Nonlinear Quantum Regime,
Phys. Rev. Lett. {\bf 111}, 133601 (2013).

\bibitem{Peng}B. Peng, S. K. ${\rm \ddot{O}}$zdemir, W. Chen, F. Nori, and L. Yang,
What is and what is not electromagnetically induced transparency in whispering-gallery microcavities,
Nat. Commun. {\bf 5}, 5082 (2014).

\bibitem{Zhangs}S. Zhang, D. A. Genov, Y. Wang, M. Liu, and X. Zhang,
Plasmon-induced transparency in metamaterials,
Phys. Rev. Lett. {\bf 101}, 047401 (2008).

\bibitem{Papa}N. Papasimakis, V. A. Fedotov,  N. I. Zheludev, and S. L. Prosvirnin,
                     Metamaterial analog of electromagnetically induced transparency,
Phys. Rev. Lett. {\bf 101}, 253903 (2008).

\bibitem{Tassin} P. Tassin, L. Zhang, T. Koschny, E. N. Economou, and C. M.  Soukoulis,
                 Low-loss metamaterials based on classical electromagnetically induced transparency,
Phys. Rev. Lett. {\bf 102}, 053901 (2009).

\bibitem{NLiu} N. Liu, L. Langguth, T. Weiss, J. K\"{a}stel, M. Fleischhauer, T. Pfau, and H. Giessen,
Plasmonic analogue of electromagnetically induced transparency at the Drude damping limit,
Nat. Mat. {\bf 8}, 758 (2009).

\bibitem{Chen0}C. Chen, I. Un, N. Tai, and T. Yen,
Asymmetric coupling between subradiant and superradiant
plasmonic resonances and its enhanced sensing performance,
Opt. Expr. {\bf 17}, 15372 (2009).

\bibitem{Dong} Z. Dong, H. Liu, J. Cao, T. Li, S. Wang, S. Zhu, and X. Zhang,
Enhanced sensing performance by the plasmonic analog of electromagnetically
induced transparency in active metamaterials,
Appl. Phys. Lett. {\bf 97}, 114101 (2010).

\bibitem{Liu} N. Liu, M. Hentschel, T. Weiss, A. P. Alivisators, and H. Giessen,
Three-dimensional plasmon rulers,
Science {\bf 332}, 1407 (2011).

\bibitem{han}Z. Han and S. I. Bozhevolnyi,
Plasmon-induced transparency with detuned ultracompact Fabry-Perot resonators in integrated
plasmonic devices,
Opt. Expr. {\bf 19}, 3251 (2011).

\bibitem{Gu}J. Gu, R. Singh, X. Liu, X. Zhang, Y. Ma, S. Zhang, S. A. Maier, Z. Tian, A. K. Azad,
H.-T. Chen, A. J. Taylor, J. Han, and W. Zhang,
Active control of electromagnetically induced transparency analogue in terahertz metamaterials,
Nat. Commun. {\bf 3}, 1151 (2012).

\bibitem{TNakanishi}T. Nakanishi, T. Otani, Y. Tamayama, and M. Kitano,
Storage of electromagnetic waves in a metamaterial that mimics
electromagnetically induced absorption in plasmonics,
Phys. Rev. B {\bf 87}, 16110(R) (2013).

\bibitem{Sun}Y. Sun, Y. Tong, C. Xue, Y. Ding, Y. Li, H. Jiang, and H. Chen,
Electromagnetic diode based on nonlinear electromagnetically induced transparency in metamaterials,
Appl. Phys. Lett. {\bf 103}, 091904 (2013).

\bibitem{Chen} J. Chen, P. Wang, C. Chen, Y. Lu, H. Ming, and Q. Zhan,
Plasmonic EIT-like switching in bright-dark-bright plasmon resonators,
Opt. Expr. {\bf 19}, 5970 (2013).
%

\bibitem{Mats}T. Matsui, M. Liu, D. A. Powell, I. V. Shadrivov, and Y. S. Kivshar,
Electromagnetic tuning of resonant transmission in magnetoelastic metamaterials,
Appl. Phys. Lett. {\bf 104}, 161117 (2014).

\bibitem{NakanishiPRApp} T. Nakanishi and M. Kitano,
Implementation of Electromagnetically Induced Transparency in a Metamaterial Controlled with Auxiliary Waves,
Phys. Rev. Appl. {\bf 4}, 024013 (2015).


\bibitem{bai} Z. Bai, G. Huang, L. Liu, and S. Zhang,
Giant Kerr nonlinearity and low-power gigahertz solitons via plasmon-induced transparency,
Sci. Rep. {\bf 5}, 13780 (2015).

\bibitem{Bai2} Z. Bai and G. Huang,
Plasmon dromions in a metamaterial via plasmon-induced transparency,
Phys. Rev. A {\bf 93}, 013818 (2016).
%

\bibitem{Bai3} Z. Bai, Datang Xu, and G. Huang,
Storage and retrieval of electromagnetic waves with orbital angular momentum via plasmon-induced transparency,
Opt. Expr. {\bf 25}, 785 (2017).


\bibitem{Lukin} M. D. Lukin, P. R. Hemmer, M. L\"{o}ffler, and M. O. Scully,
Resonant Enhancement of Parametric Process via Radiative Interference and Induced Coherence,
Phys. Rev. Lett. {\bf 81}, 2675 (1998).
%

\bibitem{Korsunsky1} E. A. Korsunsky, N. Leinfellner, A. Huss, S. Baluschev, and L. Windholz,
Phase-dependent electromagnetically induced transparency,
Phys. Rev. A {\bf 59}, 2302 (1999).

\bibitem{Korsunsky2} E. A. Korsunsky and D. V. Kosachiov,
Phase-dependent nonlinear optics with double-$\Lambda$ atoms,
Phys. Rev. A {\bf 60}, 4996 (1999).
%

\bibitem{Merriam} A. J. Merriam, S. J. Sharpe, M. Shverdin, D. Manuszak, G. Y. Yin, and S. E. Harris,
Efficient Nonlinear Frequency Conversion in an All-Resonant Double-$\Lambda$ System,
Phys. Rev. Lett. {\bf 84}, 5308 (2000).

\bibitem{DengPRA1} M. G. Payne and L. Deng,
Consequences of induced transparency in a double-$\Lambda$ scheme: Destructive interference in four-wave mixing,
Phys. Rev. A {\bf 65}, 063806 (2002).

\bibitem{Kang} H. Kang, G. Hernandez, and Y. Zhu,
Resonant four-wave mixing with slow light,
Phys. Rev. A {\bf 70}, 061804(R) (2004).

\bibitem{Ying} Y. Wu and X. Yang,
Highly efficient four-wave mixing in double-$\Lambda$ system in ultraslow propagation regime,
Phys. Rev. A {\bf 70}, 053818 (2004).
%

\bibitem{DengPRA3} L. Deng, M. G. Payne, G. Huang, and E. W. Hagley,
Formation and propagation of matched and coupled ultraslow optical soliton pairs in a four-level double-$\Lambda$ system,
Phys. Rev. E {\bf 72}, 055601(R) (2005).

\bibitem{YingSoliton} Y. Wu,
Two-color ultraslow optical solitons via four-wave mixing in cold-atom media,
Phys. Rev. A {\bf 71}, 053820 (2005).

\bibitem{Kang1} H. Kang, G. Hernandez, J. Zhang, and Y. Zhu,
Phase-controlled light switching at low light levels,
Phys. Rev. A {\bf 73}, 011802R (2006).

\bibitem{IteYu} Z.-Y. Liu, Y.-H. Chen, Y.-C. Chen, H.-Y. Lo, P.-J. Tsai, I. Yu, Y.-C. Chen, and Y.-F. Chen,
Large Cross-Phase Modulations at the Few-Photon Level,
Phys. Rev. Lett. {\bf 117}, 203601 (2016).


\bibitem{Petrosyan} D. Petrosyan and Y. P. Malakyan,
Magneto-optical rotation and cross-phase modulation via coherently driven four-level atoms in a tripod configuration,
Phys. Rev. A {\bf 70}, 023822 (2004).

\bibitem{Rebic}S. Rebi\'{c}, D. Vitali, C. Ottaviani, P. Tombesi, M. Artoni, F. Cataliotti, and R. Corbal\'{a}n,
Polarization phase gate with a tripod atomic system,
Phys. Rev. A {\bf 70}, 032317 (2004).

\bibitem{Beck}S. Beck and I. E. Mazets,
Propagation of coupled dark-state polaritons and storage of light in a tripod medium,
Phys. Rev. A {\bf 95}, 013818 (2017).



\bibitem{Joshi}A. Joshi and M. Xiao,
Generalized dark-state polaritons for photon memory in multilevel atomic media,
Phys. Rev. A {\bf 71}, 041801 (2005).

\bibitem{GaoJY}J.-Y. Gao, S.-H. Yang, D. Wang, X.-Z. Guo, K.-X. Chen, Y. Jiang, and B. Zhao,
Electromagnetically induced inhibition of two-photon absorption in sodium vapor,
Phys. Rev. A {\bf 61}, 023401 (2000).

\bibitem{Kha}U. Khadka, Y. Zhang, and Min Xiao,
Control of multitransparency windows via d ark-state phase manipulation,
Phys. Rev. A {\bf 81}, 023830 (2010).

\bibitem{LiuYM}Y.-M. Liu, X.-D. Tian, D. Yan, Y. Zhang, C.-L. Cui, and J.-H. Wu,
Nonlinear modifications of photon correlations via controlled single and double Rydberg blockade,
Phys. Rev. A {\bf 91}, 043802 (2015).


\bibitem{Ott}C. Ottaviani, D. Vitali, M. Artoni, F. Cataliotti, and P. Tombesi,
Polarization Qubit Phase Gate in Driven Atomic Media,
Phys. Rev. Lett. {\bf 90}, 197902 (2003).

\bibitem{Mat}A. B. Matsko, I. Novikova, G. R. Welch, and M. S. Zubairy,
Enhancement of Kerr nonlinearity by multiphoton coherence,
Opt. Lett. {\bf 28}, 96 (2003).

\bibitem{Rebic1}S. Rebi\'{c}, C. Ottaviani, G. Di Giuseppe, D. Vitali, and P. Tombesi,
Assessment of a quantum phase-gate operation based on nonlinear optics,
Phys. Rev. A {\bf 74}, 032301 (2006).

\bibitem{HangC}C. Hang and G. Huang,
Weak-light ultraslow vector solitons via electromagnetically induced transparency,
Phys. Rev. A {\bf 77}, 033830 (2008).


\bibitem{Rus}J. Ruseckas, V. Kudria\u{s}ov, I. A. Yu, and G. Juzeli\={u}nas,
Transfer of orbital angular momentum of light using two-component slow light,
Phys. Rev. A {\bf 87}, 053840 (2013).

\bibitem{Lee}M.-J. Lee, J. Ruseckas, C.-Y. Lee, V. Kudrias\u{s}ov, K.-F. Chang,
H.-W. Cho, G. Juzeli\={u}nas, and  I. A. Yu,
Experimental demonstration of spinor slow light,
Nat. Commun. {\bf 5}, 5542 (2014).

\bibitem{note00}Bright state (dark state) is an eigenstate of the Hamiltonian that involves (does not involve) the upper states $|3\rangle$ and $|4\rangle$.

\bibitem{note01}In quantum mechanics, a two-level atom is equivalent to an oscillator. The double-$\Lambda$-type atom has four levels, and hence is equivalent to three oscillators.

\bibitem{note02}The frequency and wave number of $l$th probe field are given by
$\omega_{pl}+\omega$ and $k_{pl}+K_a(\omega)$ ($l=1,2$), respectively. Thus $\omega=0$ corresponds to the central frequency of the probe field.

\bibitem{note1000}A similar model was also considered in Ref.~\cite{Chen}, but in which a different frequency region was chosen and no atomic FWM analogue and no study of nonlinear excitations were given.

\bibitem{Rb87} D. Steck, $^{87}$Rb D Line Data, http://steck.us/alkalidata.


\bibitem{note03} For the chosen geometry and parameters of the PIT-based metamaterial shown in Fig.~\ref{Fig2}, the resonance frequencies of the CWs and the SRR are approximately equal, and the damping rate of the dark oscillator (i.e. $\gamma_{3}$) is much smaller than those of the bright oscillators (i.e. $\gamma_{1}$ and $\gamma_{2}$); see Appendix~\ref{app2}.

\bibitem{Mikhail} M. A. Kats, N. Yu, P. Genevet, Z. Gaburro, and F. Capasso,
Effect of radiation damping on the spectral response of plasmonic components,
Opt. Expr. {\bf 19}, 21748 (2011).
%

\bibitem{LiHJ}H.-j. Li and G. Huang, Highly efficient four-wave mixing in a coherent six-level system
in ultraslow propagation regime, Phys. Rev. A {\bf 76}, 043809 (2007).

\bibitem{Lapine} M. Lapine, I. V. Shadrivov, and Y. S. Kivshar,
{\it Colloquium}: Nonlinear metamaterials,
Rev. Mod. Phys. {\bf 86}, 1093 (2014).
%

\bibitem{Shadrivov} I. V. Shadrivov, S. K. Morrison, and Y. S. Kivshar,
Tunable split-ring resonators for nonlinear negative-index metamaterials
Opt. Expr. {\bf 14}, 9344 (2006).
%

\bibitem{Wang} B. Wang, J. F. Zhou, T. Koschny, and C. M. Soukoulis,
Nonlinear properties of split-ring resonators,
Opt. Expr. {\bf 16}, 16058 (2008).

\bibitem{Jeffery} A. Jeffery and T. Kawahawa,
{\it Asymptotic Method in Nonlinear Wave Theory}
(Pitman, London, 1982).


\bibitem{Newell} A. C. Newell and J. V. Moloney, {\it Nonlinear Optics}
(Addison-Wesley, Redwood City, 1992).

%

\bibitem{Hirota} R. Hirota,
{\it The direct method in soliton theory}
(Cambridge University Press, Cambridge, 2004).
%

\bibitem{Dromion} K. Nishinari and T. Yajima,
Numerical analyses of the collision of localized structures in the Davey-Stewartson equations,
Phys. Rev. E {\bf 51}, 4986 (1995).
%

\bibitem{Pol_Sel1} J. Shao, J. Li, Y.-H. Wang, J.-Q. Li, Q. Chen, and Z.-G. Dong,
Polarization conversions of linearly and circularly polarized lights through a plasmon-induced transparent metasurface,
Appl. Phys. Lett. {\bf 115}, 243503 (2014).
%

\bibitem{Pol_Sel2} C. Pelzman and S.-Y. Cho
Polarization-selective optical transmission through a plasmonic metasurface,
Appl. Phys. Lett. {\bf 106}, 251101 (2015).
%

\bibitem{Jiang} T. Jiang, K. Chang, L. Si, L. Ran, and H. Xin,
Active microwave negative-index metamaterial transmission line with gain,
Phys. Rev. Lett. {\bf 107}, 205503 (2011).
%

\bibitem{Ye} D. Ye, K. Chang, L. Ran, and H. Xin,
Microwave gain medium with negative refractive index,
Nat. Commun. {\bf 5}, 5841 (2014).
%

\end{references}
\end{document}